\documentclass[cpp,a4paper,fleqn]{w-art}

\usepackage{times,cite,w-thm}
\theoremstyle{plain}

\theoremstyle{definition}

\usepackage[]{graphicx}

\usepackage{subfigure}
\usepackage[english]{babel}

\begin{document}

\DOIsuffix{theDOIsuffix}

\Volume{XX}
\Month{XX}
\Year{XXXX}

\pagespan{1}{}

\Receiveddate{XXXX}
\Reviseddate{XXXX}
\Accepteddate{XXXX}
\Dateposted{XXXX}


\title[Structure of Yukawa balls]{Structure and Phase transitions of Yukawa balls}
\author[H. Baumgartner]{Henning Baumgartner\inst{1}
  \footnote{Corresponding author\quad E-mail:~\textsf{baum@theo-physik.uni-kiel.de},
            Phone: +\,49\,431\,880-4063,
            Fax: +\,49\,431\,880-4094}}
\address[\inst{1}]{Institute for Theoretical Physics and Astrophysics, Christian Albrechts University Kiel, \\ Leibnizstrasse 15, D-24098 Kiel, Germany}
\author[D. Block]{Dietmar Block\inst{2}}
\address[\inst{2}]{Institute for Experimental and Applied Physics, Christian Albrechts University Kiel,\\ Leibnizstrasse 19, D-24098 Kiel, Germany}
\author[M. Bonitz]{Michael Bonitz\inst{1}}

\begin{abstract}
	In this review, an overview of structural properties and phase transitions in finite spherical dusty (complex) plasma crystals -- so-called Yukawa balls -- is given. These novel kinds of Wigner crystals 
can be directly analyzed experimentally with video cameras. The experiments clearly reveal a shell structure and 
allow to determine the shell populations, to observe metastable states and transitions between configurations as well as phase transitions. The experimental observations of the static properties are well explained by a rather simple theoretical model which treats the dust particles as being confined by a parabolic potential and interacting via an isotropic Yukawa pair potential. The excitation properties of the Yukawa balls such as normal modes and the dynamic behavior, including the time-dependent formation of the crystal requires, in addition, to include the effect of friction between the dust particles and the neutral gas. Aside from first-principle molecular dynamics and Monte Carlo simulations several analytical approaches are reviewed which include shell models and a continuum theory.
A summary of recent results and theory-experiment comparisons is given and questions for future research activities are outlined.
\end{abstract}

\maketitle

\section{Introduction}
Complex plasmas are gas plasmas consisting of electrons, ions, and neutral atoms that additionally contain microscopic particles with sizes ranging from $10$nm to some $10\mu$m. This state of matter is ubiquitous in space, e.g. in the interplanetary medium, in interstellar clouds, in comet tails, and in the ring systems of the giant planets as well as in mesospheric noctilucent clouds \cite{goertz89,northrop92}. At the same time, in microchip manufacturing, avoiding particle contamination during the many production steps that involve plasmas is a technological challenge \cite{bennett89,wu90}. On the other hand, the growth, transport, and deposition of nanoparticles is the central goal of many plasma deposition techniques, e.g. in the manufacturing of amorphous solar cells \cite{laure91,boeuf92,hollenstein00,howling00}. 

Since the early 1990s, research in complex plasmas is rapidly evolving. The original name dusty plasmas is nowadays often replaced by ``complex plasmas'' -- in analogy to complex fluids, in order to emphasize many fundamentally different properties of this medium that distinguish it from ordinary gas plasmas. The field of complex plasmas is now maturing and the interested reader can find timely monographs, e.g. on dusty plasmas in space \cite{yaroshenko95}, technological applications \cite{bouchoule99}, or on waves in dusty space plasmas \cite{verheest00}. The microparticles in complex plasmas are electrically charged by collection of plasma electrons and ions as well as by photoemission or secondary electron emission \cite{goertz89,northrop92}. In laboratory plasmas usually the collection processes dominate and the particles attain a high negative charge of a few thousand elementary charges. 
To some extent, complex plasmas behave like negative ion plasmas. In both 
cases a heavy, negatively charged species substitutes part of the electrons 
and thus affects the mobility and the contribution to shielding by negative charges. 

Most ordinary plasmas in space and laboratory are weakly coupled, which means that the interaction energy of nearest neighbours is much smaller than their thermal energy. This situation is completely different in complex plasmas with micrometer sized particles which then become strongly coupled \cite{morfill02}: the negative charge can be so large that the interaction energy of nearest neighbours exceeds their thermal energy by several orders of magnitude. A consequence of strong coupling is the formation of liquid or solid phases. In 1934, a liquid to solid phase transition had been predicted by Eugene {\sc Wigner} for the three-dimensional (3D) electron gas in metals at low temperature and low density due to strong Coulomb repulsion \cite{wigner34}. This phase transition became known as Wigner crystallization, and the solid phase as Wigner or Coulomb crystals. This kind of crystallization has been studied for decades in a varity of systems ranging from electrons trapped on top of liquid helium \cite{adams79,rousseau09}, electrons trapped in semiconductor quantum wells \cite{etienne88} and quantum dots \cite{filinov01}, strongly coupled radio-frequency discharges of dusty plasmas \cite{lin94}, hole crystallization in semiconductors \cite{bonitz06_2} to laser-cooled trapped ions in Paul- and Penning traps \cite{wineland87,drewsen98}. 
It is remarkable that these physically completely different systems exhibit very similar collective behavior which is due to the governing role of the long-range Coulomb interaction. 

What makes the dust systems particularly attractive is that here collective phenomena, such as crystallization, occur at a comparatively large length scale of several millimeters and at room temperature which makes detailed experimental investigations rather simple. 
The formation of ordered solid phases in dusty plasmas was predicted by {\sc Ikezi} \cite{ikezi86} and first observed as plasma crystals in the early 1990s \cite{li94,moehlmann94,tachibana94}. Since then crystalline structures and dynamical processes have been observed in a variety of confinement geometries ranging from 1D to 3D, see e.g. Ref.~\cite{piel02} for an overview and additional references.  The present review will focus on the 3-dimensional spherical dust crystals recently observed by {\sc Arp} et al. \cite{arp04}. Their 
 crystalline structure was found to be very similar to those observed before in ion crystals in Paul traps 
and the crystals which have been predicted to exist in expanding neutral plasmas \cite{pohl04,killian04}.
Compared to the latter, the advantage of these so called  Yukawa balls is that the individual particle positions and trajectories can 
be directly visually observed and traced and recorded with video cameras. As a result, very detailed comparisons to theory and computer simulations can be performed which go far beyond comparisons in more traditional plasmas. In particular, the comparison is not limited to average quantities such as distribution functions on long time scales, but also covers correlation effects, fluctuations, and even 
short-time effects become accessible.

Thus, the aim of this review is to show, on the example of Yukawa balls, that 
 complex plasmas are an excellent object to study in great detail static and dynamic phenomena 
 in finite systems such as ground and metastable states, transitions between 
 configurations and phase transitions. These properties have, in the past few years, been studied in a very close connection between experiment, theory and computer simulation. 
For this purpose the review is organized 
 as follows: In Section \ref{s:experiment} we discuss the experimental realization of the Yukawa balls. Then, in Section \ref{s:theory} we give an introduction to 
the theory and computer simulations and discuss the low-temperature structural properties of Yukawa balls. Section \ref{s:transitions} is devoted to finite temperature excitations of the crystals, to structural transitions and 
the melting behavior. We conclude this paper with a discussion of the results and an outline of further research directions in Sec.~\ref{s:dis}.

\section{Experimental Realization of Yukawa balls}\label{s:experiment}

To realize and investigate spherical $3$D dust clouds two ingredients are indispensable: an isotropic confinement 
to prevent Coulomb explosion of the alike charged particles and powerful 3D diagnostics. For a detailed report on the progress in the field of 3D diagnostic tools we refer the reader to \cite{block_hoboken}. Here, we will limit the discussion to dust confinement, because the shape of the confinement potential is a central input parameter for theory and simulation.

To confine dust particles inside a laboratory plasma reactor, the gravitational force has to be balanced. Thus, dust 
confinement is typically achieved in the plasma sheath region above an electrode where strong electric fields are present \cite{liebermann94a}. The trapping of the dust particles in horizontal direction is established by depressions of the electrode surface \cite{pieper96a} or flat metal rings on the electrode \cite{trottenberg95a}. However, this results in a very anisotropic confinement potential. 
The confinement in vertical direction is much stronger and thus these dust clouds are mostly 2D systems which nevertheless can form highly ordered crystals with a hexagonal lattice structure \cite{li94,tachibana94,melzer94a,moehlmann94}. Further, the supersonic ion flow towards the electrode is focused below each particle \cite{melandso95a,melandso96b,melzer96b,nambu95a,schweigert96a}. In multilayer systems, the resulting positive space-charge attracts particles in a lower layer. This process is responsible for chain formation observed in all dust clouds which are confined in regions of strong electric fields, i.e.\ regions with strong ion flows. This alignment vanishes if small particles and high gas pressure are used \cite{hayashi99a,pieper96c,zuzic00a}. However, extended homogeneous 3D plasma crystals cannot be generated this way and investigations of dust dynamics are not feasible due to strong damping.

To produce extended 3D dust clouds, various approaches were followed. {\sc Merlino} and coworkers confined dust
in a magnetized anodic plasma \cite{Barkan95c,thompson97} and investigated dust acoustic waves. Later, similar experiments on dust dynamics were performed in other discharges \cite{Fortov00a,ratynskaia04a,ratynskaia04b,Thomas01b,trottenberg06,pilch07,pilch08}. To
produce 3D plasma crystals, a number of experiments have been performed under microgravity conditions. These experiments have provided many interesting observations, e.g.\ of localized crystalline structures \cite{nefedov03}, 
of complex plasma boundaries \cite{annaratone03,bryant04a}, of coalescence of complex plasma fluids \cite{ivlev06a}, 
of transport properties \cite{fortov04f,fortov03c}, and of low-frequency waves and instabilities
\cite{khrapak03a,piel06c,samsonov03b,yaroshenko04c}. However, the most striking observation was the formation of a 
dust-free zone (void) in the center of the discharge \cite{morfill99c,nefedov03}. It was proposed that the ion drag
force is responsible for the formation of voids \cite{Goree99a,tsytovich01a,tsytovich01d,tsytovich01e}, and a
 combination of simulations \cite{akdim02a,land05c,land06a}, experiments \cite{klindworth04a,praburam96,Rothermel02a,wolter07}, 
and recent ion drag models \cite{Hutchinson05a,Hutchinson06a,khrapak02} were able to verify this.
Although it was shown very recently that a void closure can be achieved \cite{lipaev07}, the formation of void-free
crystalline dust clouds is still an important issue. Such dust clouds in a well defined confinement would allow
to extend a large fraction of the successful research on 2D plasma crystals to 3D. Interesting observations were reported by {\sc Annaratone} and coworkers \cite{annaratone04,antonova06a} who observed spherical dust clouds with less than $50$ particles in a secondary discharge in front of an adaptive electrode. Unfortunately, these clouds are rather in a liquid state and their confinement is not yet understood.

\begin{figure}[t]\label{fig:yb_exp_setup}
\centering
\includegraphics[width=0.95\textwidth]{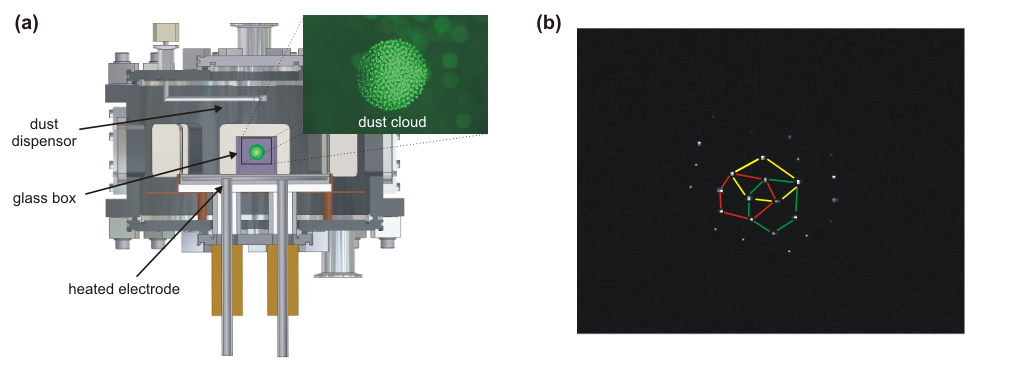}
\caption{(a) Side view of the discharge arrangement for Yukawa balls. The lower electrode is heated ($T<90^\circ$C). The vacuum vessel is grounded and kept at room temperature. The dust cloud is confined inside a glass cube where the upper and lower side are left open. The inset shows an image of a large dust cloud with $1\,$cm in diameter. (b) A thin slice at the front side of the cloud is illuminated. The particles basically arrange in a hexagonal lattice. After Ref.~\cite{arp04}.}
\end{figure}

A different approach to confine void-free dust clouds is motivated by experiments, simulations and theories on void formation \cite{Akdim04a,Goree99a,morfill99c,Rothermel02a}. Those results showed that a void-free dust confinement at moderate plasma densities is hampered by the inherent ion flows towards the discharge boundaries. As a consequence, {\sc Arp} and coworkers \cite{arp04} have modified the usual setup of an asymmetric capacitively coupled rf-discharge [Fig.~\ref{fig:yb_exp_setup}(a)]. To establish a vertical temperature gradient the rf-electrode was heated to $T=(60-80)^\circ$C. The grounded vacuum vessel was kept at room temperature. The resulting thermophoretic force should balance gravity for particles with a radius $r_p<3\,\mu$m, i.e. allow for dust levitation in the bulk plasma. In addition to the experimental setup used by {\sc Rothermel} \cite{Rothermel02a}, 
Arp et al. proposed to place a glass box with square cross section on the electrode that was left open at the top and at the bottom. For low discharge power ($P<10\,$W), they discovered that dust, which was dropped into the glass box, formed spherical void-free clouds \cite{arp04}. A picture of a laser illuminated dust cloud (Yukawa ball) consisting of roughly $10,000$ dust particles is shown in the inset of Fig.~\ref{fig:yb_exp_setup}(a). Already if only the front of such a dust cloud was illuminated 
with a laser sheet, interesting structural properties were observed [Fig.~\ref{fig:yb_exp_setup}(b)]. The particle arrangement was static and particles on the surface of Yukawa balls seemed to have preferably $5$ or $6$ nearest neighbors. Moreover, the particles did not form vertical chains, i.e.\  first evidence was found that the Yukawa balls were in a solid or even crystalline state. 

\begin{figure}[t]\centering
\includegraphics[width=0.9\textwidth]{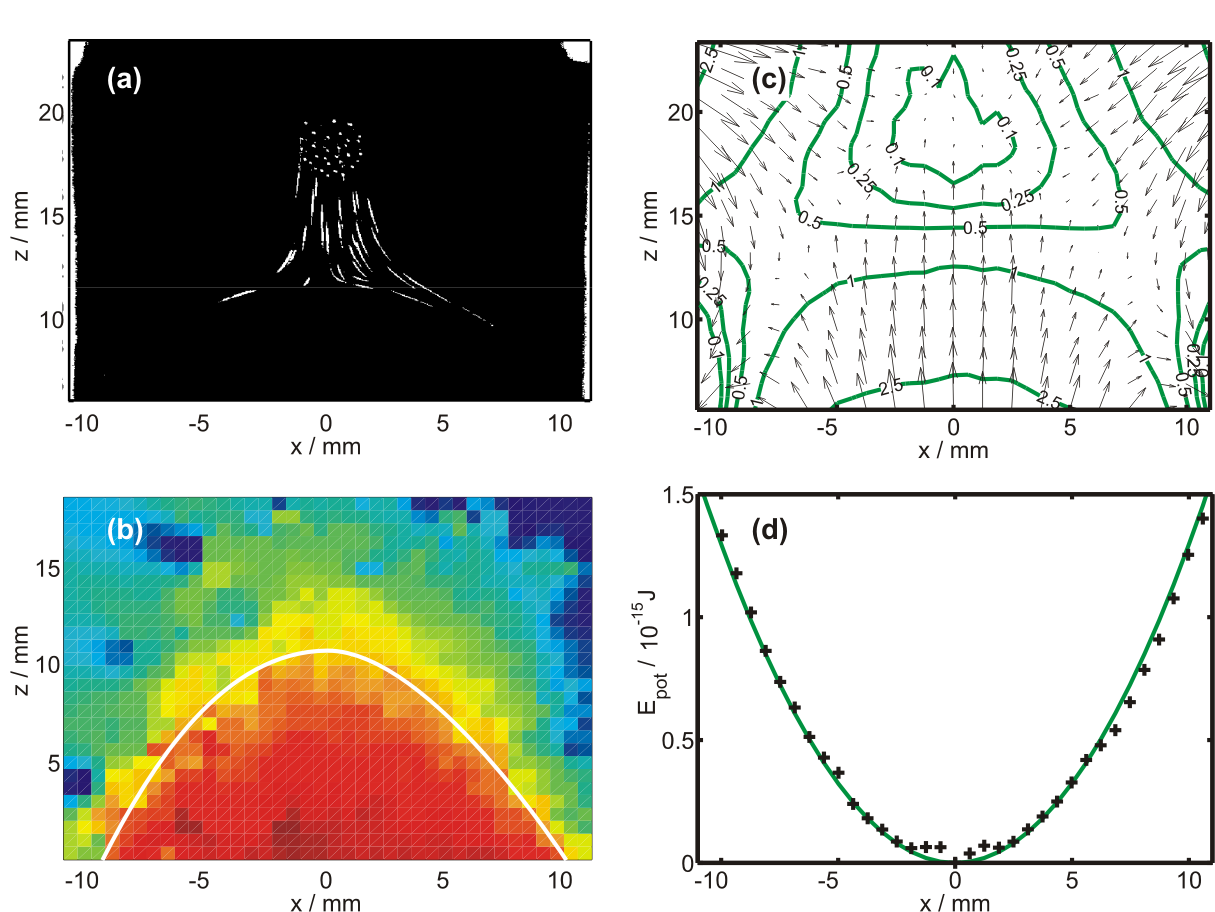}
\caption{Experiments on the confinement of Yukawa balls: 
(a) Superposition of multiple video frames. As soon as the discharge is switched off, the trapped particles fall down. Their motion is only affected by gravity and the thermophoretic force. 
(b) Vertical component of the experimentally determined thermophoretic force field. In 'red' regions the thermophoretic force exceeds gravity, in 'blue' regions gravity is dominant. The solid line shows where both forces balance. 
(c) Trap potential obtained from PIV measurements and fluid simulations. 
(d) A horizontal section through the trap center reveals an almost parabolic confinement. From Ref.~\cite{arp2006a}.}
\label{fig:yb_confinement}
\end{figure}

To achieve a quantitative description of the dust confinement, fluid simulations with the {\sc Siglo-2D} code \cite{boeuf95,boeuf94b} in combination with additional experiments were performed \cite{arp2006a}. The results are summarized in Fig.~\ref{fig:yb_confinement}. It was found that thermophoresis alone was not sufficient to explain the confinement [Fig.~\ref{fig:yb_confinement}(a)]. By means of particle imaging velocimetry (PIV) experiments the thermophoretic force field was measured [Fig.~\ref{fig:yb_confinement}(b)]. It was found that thermophoresis contributed about $70$ percent to the vertical confinement and that the radial confinement and the  remaining $30$ percent in vertical direction were provided by plasma induced forces, namely electrostatic fields due to surface charges on the glass \cite{arp2006a}. Further, the fluid simulations showed that the Yukawa balls were confined in a region where the plasma production was negligible. Due to the low plasma density ($n \approx 10^{13} {\rm m}^{-3}$) and the weak electric fields the ion drifts were significantly below the ion sound speed. The resulting ion drag force was about two orders of magnitude smaller than the electrostatic confinement force and could, therefore, be neglected and chain formation is not expected. The  confinement provided by the 
combination of gravity, thermophoresis and electric fields is plotted in Fig.~\ref{fig:yb_confinement}(c) and is in very good agreement with the experimental observations [Fig.~\ref{fig:yb_confinement}(a)]. It yields the correct levitation height, its depth is sufficient to trap particles with a few thousand elementary charges and even its asymmetry was observable for huge clouds in the experiment \cite{arp2006a}. 

For further structural analysis and especially for comparison with simulations as well as with other strongly coupled systems, it is important to note that the trapping potential is essentially isotropic and parabolic for dust clouds with a radius of less than $2\,$mm, i.e.\ $N<1000$ particles. Furthermore, for the trapping process an interaction of dust and plasma, as proposed by {\sc Totsuji} \cite{totsuji05a,totsuji05c}, is not required. The particle trap is provided by external forces only and thus this trap geometry is closely related to those of Penning and Paul traps \cite{paul90}. This allows for a direct comparison of the structural properties of Yukawa balls and e.g. trapped laser-cooled ions.

\section{Theoretical Description of Yukawa balls}\label{s:theory}
A theoretical description of complex plasmas is very complicated, due to the large number of different particle species, which include electrons, ions, neutrals and dust particles, and their huge difference in mass. Thus the length and time scales of the motion of the different particles differ by many orders of magnitude, making a selfconsistent modeling unfeasible. The most advanced computer models use particle in cell (PIC) simulations with a full account of the time-dependence of the discharge.\cite{matyash07}. However, this currently allows to treat only a very small number of dust particles in the plasma environment. A simplified model which treats the streaming electrons and ions within linear response and computes the dynamically screened potential of the dust particles was developed by {\sc Joyce} and {\sc Lampe}, see \cite{joyce1} and references therein. Both simulations yield an anisotropic and non-monotonic (wake) potential around a dust grain if the ion streaming velocity exceeds the sound speed which is normally the case in the plasma sheath. At the same time, simulations for the conditions of Yukawa balls which are produced in the plasma bulk revealed that there the streaming velocities are much lower and the dust-dust interaction is nearly isotropic and statically screened. It is, therefore, well approximated by a Yukawa potential. These findings justify the use of a much simpler theoretical model which is discussed in the following sections.

\subsection{Theoretical Model}\label{ss:model}
A standard model to describe the Yukawa balls and their dynamics is based on the Hamiltonian
\begin{equation}
\label{equation1}
	H (\textbf{r}_1\dots , \textbf{p}_1, \dots \textbf{r}_N, \dots \textbf{p}_N,t) = 
\sum_{i=1}^N \frac{\textbf{p}_i^2}{2m}  + 
\sum_{i=1}^N \frac{\alpha}{2} \textbf{r}_i^2 + 
\sum_{i=1}^N U^{\rm ext}(\textbf{r}_i,t) + 
\sum_{i\ne j}^N \frac{q^2}{4 \pi \epsilon  r_{ij}} \cdot e^{-\kappa r_{ij}},
\end{equation}
containing kinetic energy, a confinement and the interaction energy. The isotropic and parabolic confinement is based on the experimental results, cf. Sec.~\ref{s:experiment}. $q$ is the particle charge and $\alpha$ the confinement strength. For both values the experiment sets rather narrow boundaries \cite{arp2006a}.
Finally, $\epsilon$ is the background dielectric constant and $r_{ij}$ is the distance between particles $i$ and $j$. The effect of the surrounding plasma on the dust is condensed in the static screening parameter $\kappa$ which will be treated as a parameter below. A second important effect of the medium on the dust particles 
can be described by a friction force. It is beyond the hamiltonian model but can be included in the equations of motion, see Sec.~\ref{ss:simulations}. In writing the formula (\ref{equation1}) we assumed that all particles 
have identical mass and charge which is a good approximation of the experimental conditions. In fact, commercially available dust particles have a diameter which fluctuates by less than one percent, and the remaining charge variations were shown to  have a negligable effect on the structure of Yukawa balls \cite{golubnychiy06}.
Thus, the present model is expected to be well appropriate to reproduce the experimental conditions.

In Eq.~(\ref{equation1}) we have also included a possibly time-dependent potential $U^{\rm ext}$ which describes external manipulations of the dust particles.
The Hamiltonian (\ref{equation1}) is brought to a simpler dimensionless form suitable for numerical analysis by introducing units of length and energy, $r_0 = (q^2/16 \pi \epsilon \alpha)^{1/3}$ and $E_0 = (\alpha q^4/32 \pi^2 \epsilon^2)^{1/3}$, where $r_0$ corresponds to the ground state distance of two particles in the trap in the absence of screening and $E_0$ is the corresponding pair interaction energy. As the third scale parameter we introduce the unit of time $T_0$ which is the inverse of the trap frequency $T_0^{-1}=\omega=\sqrt{\alpha/m}$.

\subsection{Computer simulations}\label{ss:simulations}
The model (\ref{equation1}) is well suited for accurate computer simulations. Since the dust particles consitute a classical system, exact simulation methods are available which include molecular dynamics (MD), Langevin dynamics (LMD) and Metropolis Monte Carlo (MC) which have been successfully applied in recent years. Since most of the theoretical results in this paper are based on these methods we give a brief discussion.

{\em Molecular dynamics} solves Newton's equations corresponding to the hamiltonian (\ref{equation1}),
\begin{eqnarray} \label{eq:md}
 \dot{\textbf{p}}_i(t) &=& -\alpha \textbf{r}_i - {\vec \nabla} U^{\rm ext}(\textbf{r}_i,t) 
+ \sum_{j\ne i} \frac{q^2}{4 \pi \epsilon  r^3_{ij}}(1+\kappa r_{ij})\textbf{r}_{ij} \cdot e^{-\kappa r_{ij}}, 
\quad i=1,\dots N,
\\\nonumber
\textbf{r}_i(0) &=& \textbf{r}^0_i, \qquad 
\textbf{p}_i(0) = \textbf{p}^0_i,
\end{eqnarray}
where the last line contains the initial conditions. The coupled system (\ref{eq:md}) of ordinary differential equations is efficiently solved for up to several thousand particles by standard techniques such as Runge-Kutta methods. The present version gives a solution in the microcanonical ensemble where the total energy $H$ is conserved. The initial conditions are chosen as close as possible to the experimental situation. This scheme has been used to obtain the ground and metastable states of the Yukawa balls, see Sec.~\ref{ss:groundstate}

To reproduce the equilibrium properties of Yukawa balls which are characterized by a given temperature it is more appropriate to perform simulations in the canonical ensemble. This can be realized by {\em Langevin MD} where, in addition to the total force ${\bf F}_i$ on particle $i$ [r.h.s. of Eq.~(\ref{eq:md})] a friction force describing the momentum loss of the particles to the neutral gas and a random fluctuating force modeling the collisions of 
the particles with the plasma medium (Brownian motion),
\begin{eqnarray} \label{eq:lmd}
 \dot{\textbf{p}}_i(t) &=& {\bf F}_i(t) - \nu m \dot{\textbf{r}}_i +{\bf y}_i(t)
\quad i=1,\dots N,
\\\nonumber
\langle \textbf{y}_i \rangle &=& 0, 
\qquad 
\langle y_{i\alpha}(t)y_{j\beta}(t') \rangle = 2 D\delta_{i,j}\delta_{\alpha,\beta}\delta(t-t'),\qquad 
\alpha, \beta = x,y,z,
\end{eqnarray}
to be supplemented by the same initial conditions as in Eq.~(\ref{eq:md}). Choosing the noise intensity according to the Einstein relation, 
$D=m k_BT \nu$, the system relaxes towards a Maxwellian velocity distribution with temperature $T$ -- tpyically the temperature of the neutral gas. This method has been used to obtain the probability of metastable states, cf. Sec.~\ref{ss:probability_ms} and for nonequilibrium dynamical simulations of Yukawa balls \cite{kaehlert08}.

Finally, a first principle method to study the thermodynamic equilibrium properties of the dust particles is {\em Monte Carlo}. In the canonical ensemble where the dust temperature $T_d$ and the particle number are assumed to be fixed, the probability of a certain configuration $s$ of the Yukawa ball with the coordinates ${\bf R}^s=\{ {\bf r}^s_1, \dots {\bf r}^s_N\}$ is defined by 
\begin{equation}\label{eq:p_canon}
 P({\bf R}^s;T_d,N,\alpha) = \frac{1}{Z}e^{-H_0({\bf R}^s)/k_BT_d}, \qquad 
\int d^3r_1\dots d^3r_NP({\bf R}^s;T_d,N;\alpha) = 1
\end{equation}
where $Z$ is the normalization constant (partition function).
Note that in contrast to thermodynamics of macroscopic systems, here the volume is not fixed but its role is taken over by the confinement strength. The exponent $H_0$ of the probability distribution $P$ is the hamiltonian (\ref{equation1}) with the momenta put equal to zero [the momentum and coordinate distributions factorize, and the former is given by a Maxwellian and does not influence the particle configurations]. Also, the effect of dissipation is not included in this description. The equilibrium state of a Yukawa ball, is obtained by maximizing the probability $P$ which is efficiently done using the Metropolis algorithm, e.g. \cite{numbook}. MC methods have been applied to find the ground and metastable states, see Sec.~\ref{ss:groundstate}, \ref{ss:probability_ms}, and finite temperature effects and melting points, Sec.~\ref{ss:finite_t}.

\section{Structure of Yukawa balls}\label{s:structure}
Let us start the analysis of Yukawa balls by considering their ground state properties. They are obtained in the zero-temperature limit where all particles settle in the minima of the total potential energy $H_0$ given by the sum of confinement and pair interactions. While zero temperature seems to contradict the experimental situation -- the experiments are performed at room temperature -- this is in fact a very good approximation. The reason is the very high charge collected on the grain surface leading to a potential energy per particle which exceeds the kinetic energy by a factor of $500$ or more. Therefore, kinetic energy has a negligible effect on the stationary properties. 
In most cases the potential energy has several minima. In this case, the deepest minimum corresponds to the classical ground state whereas the others correspond to metastable states. Due to the spherical symmetry of the hamiltonian $H_0$ the low temperature stationary states have the same symmetry: they consist of spherically symmetric concentric shells with a very small width which is in excellent agreement with the experiment. Different states differ by the radii and populations $N_k$ of the shells. The particle arrangement within the shells is highly symmetric with a combination of five-fold and six-fold symmetry (particles dominantly have five or six nearest neighbors) \cite{arp04,ludwig05}.

\subsection{Ground and metastable states}\label{ss:groundstate}
Much work has been devoted to first-principle simulations of spherical Coulomb and Yukawa crystals. This is efficiently done by MD simulations by solving Eqs.~(\ref{eq:md}), combined with ``simulated annealing''. There one starts with a random initial configuration and reduces the velocity of all particles proportionally in small steps, until they have zero kinetic energy, \cite{kirkpatrick83,ludwig05}. Alternatively, one can use LMD simulations, Eq.~(\ref{eq:lmd}), with a very small damping coefficient $\nu$, e.g. \cite{kaehlert08} or MC. The ground state for the case of pure Coulomb interaction was reported many years ago by Hasse et al. \cite{hasse91} where also earlier references are given, for larger crystals, see also \cite{schiffer02}. These data correctly describe the general trend previously observed in spherical ion crystals. However, with the first experimental results on Yukawa balls in dusty plasmas very detailed comparisons of the precise shell structure became possible. Then it turned out that most data of Ref. \cite{hasse91} refered to metastable states rather than to the ground state. The first accurate data for spherical Coulomb crystals were obtained by {\sc Ludwig} et al. \cite{ludwig05} who performed MD with simulated annealing. Extensive tables for  $N \le 160$ and some larger clusters were presented in Ref. \cite{arp05} and the  results were later confirmed by other groups, e.g. \cite{apolinario07, abdrashitov08}. 
\begin{figure}[ht]
	\centering
	\includegraphics[width=9.5cm]{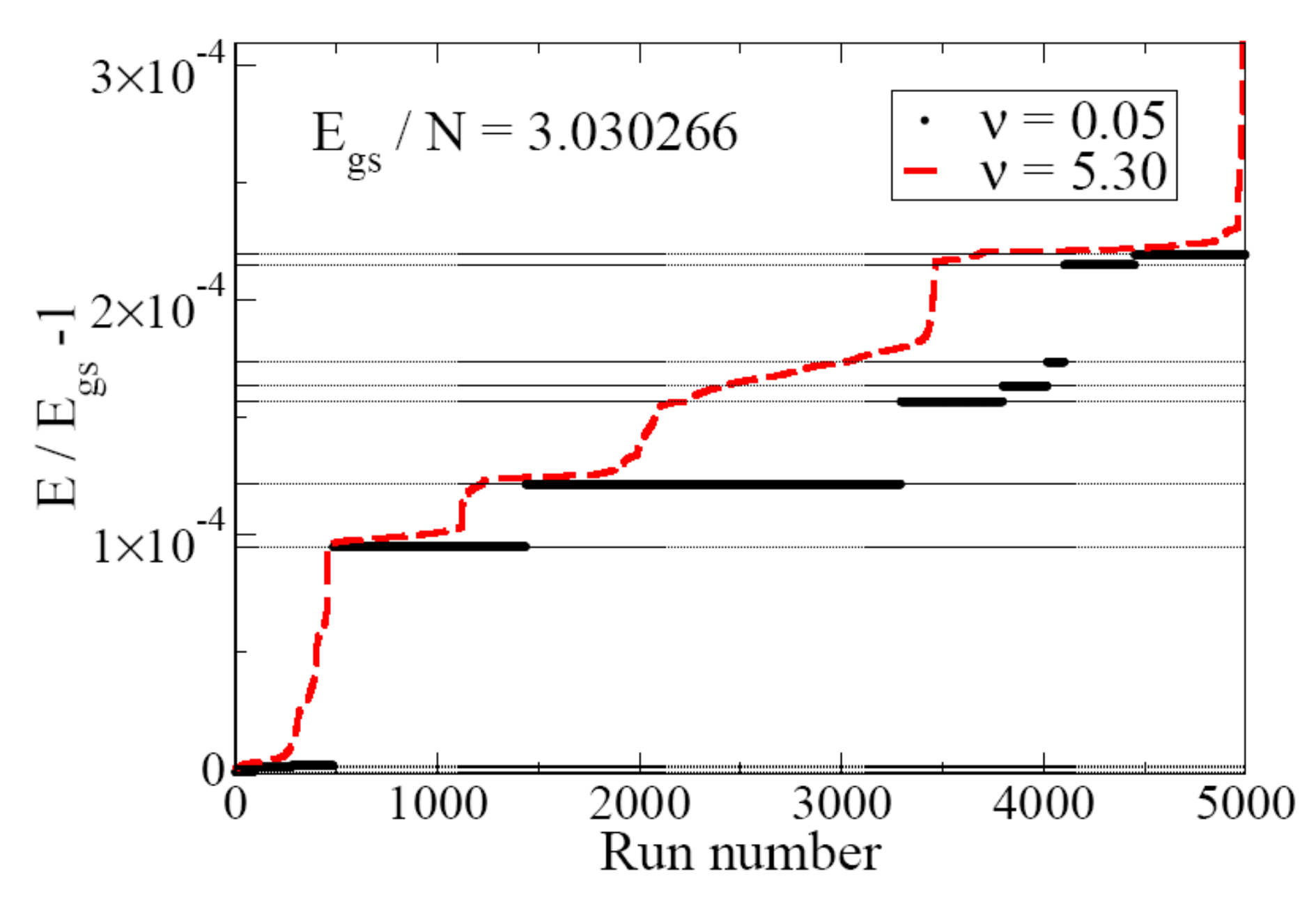}
	\caption{\label{fig2}(color online) Stationary states observed in the LMD simulations for $N = 31$, $\kappa = 1.4$ for cooling to a minimum kinetic energy $\left\langle E^{min}_{kin} \right\rangle = 10^{-8}$. The runs are sorted by the energy or the final state, see also table \ref{tab2}. For slow cooling (black bars, $\nu = 0.05$) one can clearly distinguish different states, cf. the horizontal lines. The length of the bold lines is proportional to the occurrence frequency of a state in a total of $5000$ runs. In the case of strong friction (red, dashed line, $\nu = 5.3$) the particles often lose their kinetic energy before they can settle into the equilibrium positions and the fine structure cannot be resolved. From Ref.~\cite{kaehlert08}}
\end{figure}
\begin{table}[ht]
\centering
\begin{tabular}{c c|c c}
	$\Delta$ E/N & config. & $\Delta$ E/N & config. \\
	\hline
	${\bf 3.030266}$ & ${\bf (27,4)}$ & $0.000479$ & $(26,5)$\\
	$0.000006$       & $(27,4)$       & $0.000499$ & $(26,5)$\\
	$0.000009$       & $(27,4)$       & $0.000530$ & $(26,5)$\\
	$0.000291$       & $(26,5)$       & $0.000656$ & $(25,6)$\\
	$0.000372$       & $(26,5)$       & $0.000669$ & $(25,6)$\\
	\hline
\end{tabular}
\caption{\label{tab2}Energy difference between metastable states and the ground state (the ground state and its energy is given by bold numbers) as seen in Fig.~\ref{fig2}. States with the same shell configuration but different energy differ only by the arrangement of the particles on the same shell (fine structure) \cite{ludwig05}.}
\end{table}

First structural data for Yukawa balls have only recently been obtained, e.g. \cite{bonitz06,apolinario_phd}.
The first systematic analysis was performed in Ref. \cite{baumgartner08} for small crystals with $N \le 60$ 
for a range of screening parameters of $0 \le \kappa \le 5.0$. An example of the energetically lowest stationary states is given in Fig.~\ref{fig2} and in table \ref{tab2} for a cluster with $N=31$ and $\kappa = 1.4$. 
The table also contains the shell configuration of the states. As one can see in some cases there exist several states with the same configuration which differ only in the arrangement of the particles on the shells. This ``fine structure'' was first observed in Ref.~\cite{ludwig05} and revealed by a Voronoi analysis \cite{voronoi1}. It is remarkable how small the energetic spacing of these states is which underlines the high numerical accuracy needed to find the ground states. Since no method is able to reveal the ground state with one hundred percent reliability it is necessary to perform a large series of repeated calculations, as shown in Fig.~\ref{fig2}, where 5000 runs were performed. Further, the results sensitively depend on the cooling rate in the annealing or, equivalently, on the magnitude of the friction in an LMD simulation. The ground state and the metastable states can only be resolved for sufficiently small damping, i.e. for slow cooling \cite{kaehlert08}. Further note that the number of metastable states increases exponentially with $N$ which makes it very difficult to find the ground state for clusters with more than about $1000$ particles.
\begin{figure}[ht]
	\centering
	\includegraphics[width=10cm]{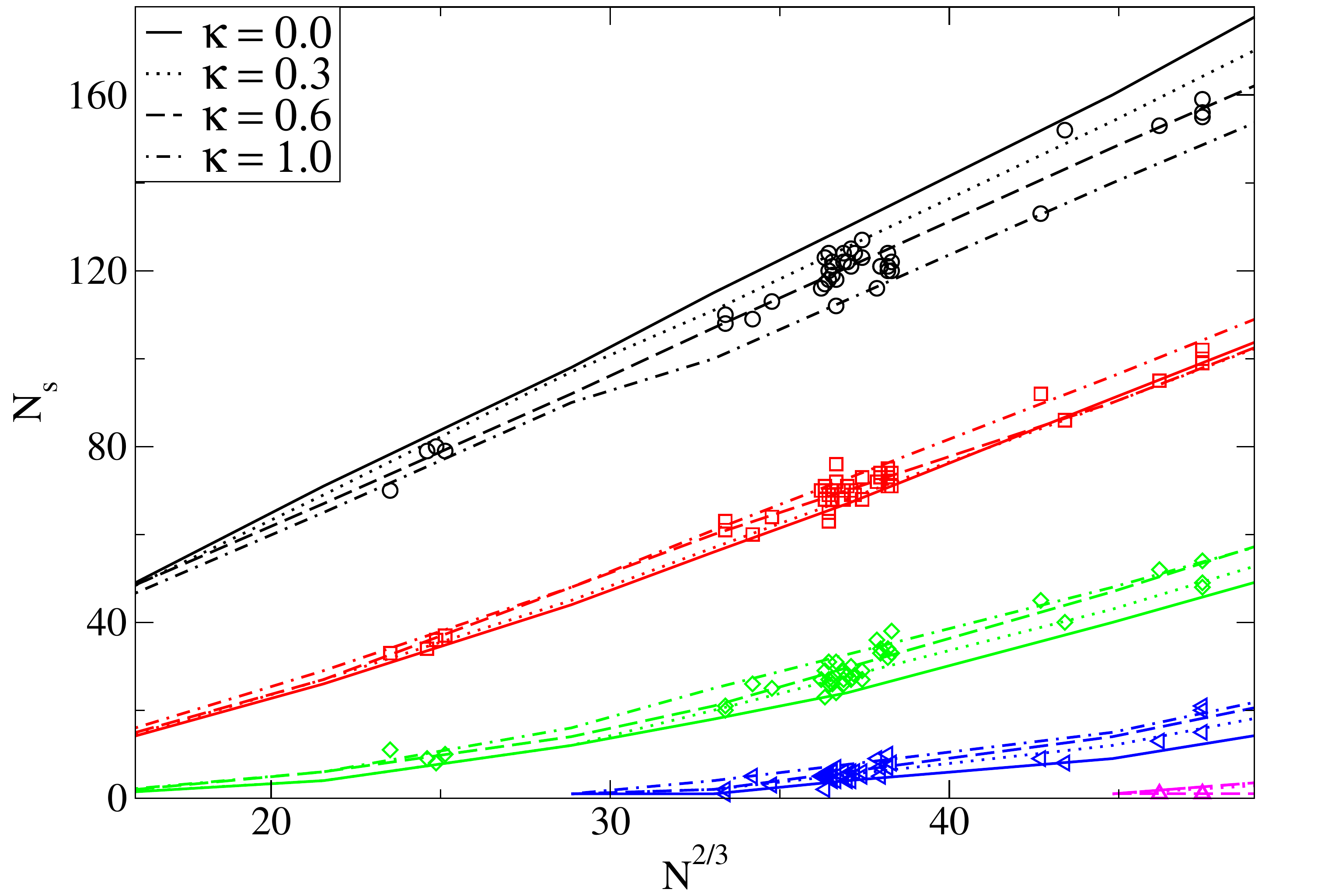}
	\caption{\label{fig3} (color online) Experimental (symbols) and theoretical (lines) shell population $N_S$ of the ground states  versus system size. The MD results are obtained for different screening parameters $\kappa$ and  show that the particles redistribute towards inner shells with increased $\kappa$. 
        From Ref.~ \cite{bonitz06}}
\end{figure}
\subsection{Effect of screening}\label{ss:screening_struc}
A comparison of experimental and theoretical cluster configurations is shown in Fig.~\ref{fig3} for particle numbers in the range of $N\le 500$. In addition to the measured configurations of $43$ clusters MD simulations of the ground states of the same particle numbers and with different values of $\kappa$ were performed \cite{bonitz06}, using simulated annealing, as described in Sec.~\ref{ss:groundstate} above. Comparison of the results shows that the model of isotropic static screening in fact reproduces the measured shell occupations very well. Furthermore, it is obvious that the measurements cannot be explained by pure Coulomb interaction. Since $\kappa$ is the only free parameter in the simulations within the model (\ref{equation1}), comparison to the measurements allows for reliable predictions of the screening strengths in the experiments which is not directly accessible. The result of $\kappa \approx 0.6$, corresponding to $\lambda_D{\bar r} = 1.5$, gives a value which agrees well with other estimates
and the accuracy was estimated to be within 50 percent \cite{bonitz06}. Sources of errors such as charge
fluctuations and finite temperature were found to be small \cite{golubnychiy06} and also the occurence of
metastable state in the experiments, see Sec.~\ref{ss:probability_ms}, has a negligible influence on this result.

For completeness, we note that Ref.~\cite{bonitz06} also provided detailed comparisons of the experimental and theoretical {\em shell radii} $R_s$. It turned out that the radii are almost equidistant and increase approximately as $N^{1/3}$, as one would expect in bulk matter. Most importantly, while the absolute value of the radii decreases with screening (due to the reduced interparticle repulsion), no effect of screening is observed for the ratio of $R_s$ to the mean interparticle distance within a shell, for fixed $N$. Since at the same time, the particle number on the outer shells decreases with $\kappa$, this has important consequences for the mean density distribution within the Yukawa balls which will be considered more in detail in Sec.~\ref{ss:continuum}.
\begin{figure}[ht]
\centering
\includegraphics[width=10cm]{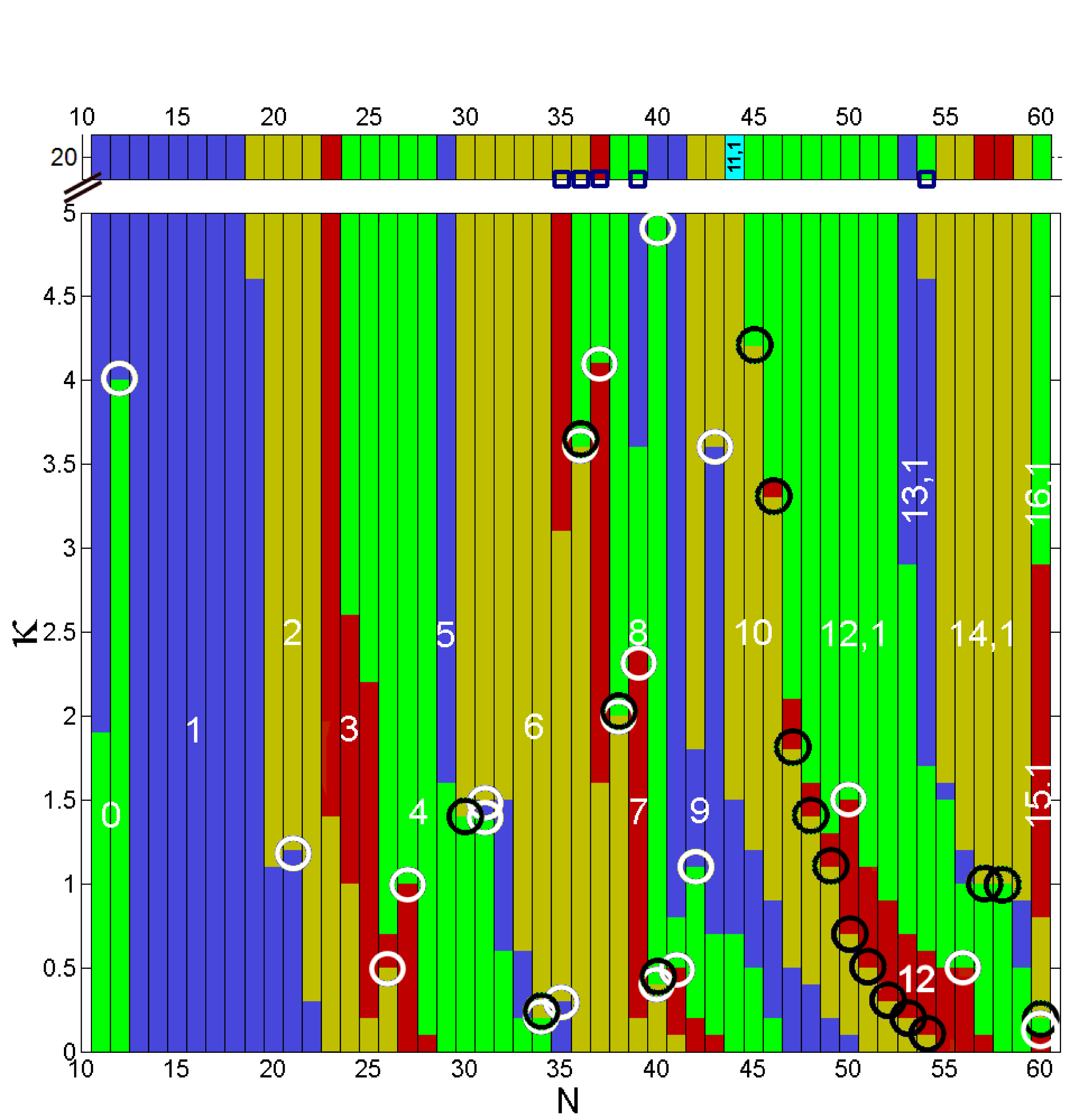}
\caption{\label{fig4}
(color online) Ground states of small Yukawa balls vs. screening parameter. The white numbers denote the number of particles on the inner shell(s). The black circles indicate anomalies of the 1st kind. The white circles indicate the end of the screening range, where anomalies of the 2nd kind appear. The ground states for a screening parameter $\kappa = 20.0$ are plotted above the diagram. The cyan bar for $N=44$ at $\kappa = 20.0$ refers to a ground state of $(11,1)$ in the center region; it is the only time this configuration is part of a ground state. The small squares just below $\kappa = 20.0$ indicate anomalies of the 3rd kind, where a ground state configuration reappears with increased screening. From Ref.~\cite{baumgartner08}}
\end{figure}	

There are two general effects of screening on the structure of Yukawa balls: first, increase of screening leads to cluster compression and, second, to increased occupation of inner shells. However, this is only the average trend, 
and noteble deviations have been found. A detailed numerical analysis of small Yukawa balls in a broad range of screening parameters has revealed a number of interesting anomalies \cite{baumgartner08}, the shell populations are shown in Fig.~\ref{fig4}. In general the particle number on the inner shell increases one by one, when $\kappa$ is increased. However, in several cases, two particles move inward simultaneously, see for example $N=30$ at $\kappa\approx 1.5$ where the configuration changes according to $(26,4)\rightarrow (24,6)$. These correlated transitions have been called  ``anomaly of the 1st kind'' and are observed 18 times. 
A second anomaly is observed when $N$ is increased by one at constant $\kappa$. The normal trend is an increase of the inner or outer shell population by one. However, in some cases the inner shell population {\em decreases} by one whereas the outer shell population increases by two. This has been observed for Coulomb balls in a single case $N=59\rightarrow 60$ \cite{tsuruta93} when the configuration changes according to $(46,12,1)\rightarrow (48,12)$. In the case of Yukawa balls this ``2nd anomaly'' is observed frequently, see e.g. $N=11\rightarrow 12$ for $2\le\kappa\le 4$ where the configuration changes according to $(10,1)\rightarrow (12,0)$. 
Finally, there are five cases ($N=35, 36, 37, 39, 54$) of 
{\em reentrant shell configuration changes} where for a fixed $N$ and increasing $\kappa$ a particle moves from an inner to an outer shell restoring a configuration which existed before, at a smaller value of $\kappa$. The reason for these non-trivial configuration changes are special symmetries of the particle arrangement which allow to reduce the total energy of the Yukawa ball. A table containing a complete list of the anomalies is given in Ref.~\cite{baumgartner08}.
	
\section{Analytical Models for the structure of Yukawa balls}\label{s:analytics}
While computer simulations provide virtually exact structural data for a certain hamiltonian a lot of additional information and physical insight can be gained from analytical models. Appropriate simplification of the system allows to observe the dominant trends and effects. Here we summarize the results of two such models -- shell models and statistical models based on a fluid-type continuum description. We start by considering the Coulomb limit which is simpler -- the ground state configurations of small clusters with $N\le 8$ can be found analytically \cite{kamimura07} -- and then proceed to Yukawa interaction.
\subsection{Shell models}\label{ss:shell_model}
The first idea of a simplified picture of the stationary particle configuration is to use a shell model, where the particles are confined to concentric shells with zero width. The first shell model was derived by {\sc Avilov} and {\sc Hasse} \cite{hasse91} and was improved by {\sc Tsuruta} and {\sc Ichimaru} \cite{tsuruta93} who included correlation effects.  {\sc Kraeft} and {\sc Bonitz} further improved this model and presented detailed comparisons with MD simulations \cite{kraeft06}. They gave the following form of the total energy per particle for a system of $L$ homogeneously charged concentric spherical shells
\begin{equation}\label{equation2}
	\frac{E_{model}(N)}{N(Ze)^2/r_0} = 2^{1/3} \sum_{\mu=1}^L \frac{N_{\nu}}{N x_{\nu}} \left ( \left [ \frac{N_{\nu}-\epsilon \sqrt{N_{\nu}}}{2} + \sum_{\mu < \nu} N_{\mu} + \zeta \right ] + \frac{1}{2} x_{\nu}^3 \right)-\frac{9}{10}N^{2/3},
\end{equation}
where $\zeta = {0,1}$ accounting for a particle in the trap center and $x_{\mu}$ is the radius of the shell $\mu$ in units of $r_0$. Here, the first term (proportional to $N^2_{\nu}$) is the surface energy of a spherical capacitor of radius $a$, containing $N_{\nu}$ charges, $E_{surf} (N) = (N(N-1)e^2)/(2a)$. The contribution from the sum over $\mu$ accounts for the electrostatic interaction of the shell $\nu$ with all inner shells. The $x^3_{\nu}$ contribution describes the confinement energy and the last term takes into account a (possible) compensating background and is not relevant for the structure of the Coulomb ball. This model, without the term proportional to $\epsilon$, can be rigorously derived from a mean field theory \cite{henning06} which is discussed in Sec.~\ref{ss:continuum} and was given by {\sc Hasse} et al. \cite{hasse91} although they missed the $-1$ correction to the surface energy $E_{surf}$. 

The cluster configuration can now be derived simply by an optimization procedure searching for those shell populations and radii which minimize the total energy (\ref{equation2}) which is much simpler than to solve the exact problem. While this yields the correct qualitative trend, however, with $\epsilon=0$ there are large quantitative deviations. {\sc Tsuruta} and {\sc Ichimaru} corrected the Hasse model by introducing a correlation correction (they used $\epsilon=1$) which reflects the discreteness of the particles and takes into account that 
the area around a particle an a shell cannot be occupied by others. Using $\epsilon=1$ allows to reproduce the shell populations within 5 percent \cite{kraeft06}. However, the exact expression for the correlation corrections is unknown, and there is no obvious reason why this coefficient should be equal to one.
In fact, a further improvement of the quality of the model is achieved by leaving $\epsilon$ as an additional free parameter that is optimized to minimze the energy. This has been done in Ref.~\cite{kraeft06} and allows to reduce the deviations from the exact ground states to $(1\dots 2)\%$. The resulting values for the correlation parameters were slightly above one and converged to $\epsilon=1.104$ for large $N$, a result which could be recently derived from the Thomson model by {\sc Cioslowski} \cite{cioslowski_08pre,cioslowski_08jcp}.

For Yukawa balls the situation is more complex. A mean field-type model ($\epsilon=0$) was recently obtained by 
{\sc Totsuji} et al. \cite{totsuji05c} and derived from continuum theory by {\sc Henning} et al. \cite{henning06},
\begin{eqnarray}
&& E_{model}(N;\kappa)=\sum_{\nu=1}^L N_\nu\biggl\{\frac{\alpha}{2}R_\nu^2+Q^2
\frac{e^{-\kappa R_\nu}}{R_\nu} \times
\nonumber\\
&&\quad \biggl(\frac{\sinh(\kappa R_\nu)}{\kappa R_\nu}\frac{N_\nu-\epsilon_{\nu}(N,\kappa)\sqrt{N_{\nu}}}{2}+\zeta+\sum_{\mu<\nu}\frac{\sinh(\kappa R_\mu)}{\kappa R_\mu}N_\mu\biggr)\biggr\},
\label{eq:ns2}
\end{eqnarray}
where, compared to Eq.~(\ref{equation2}), the background term (last term) has been dropped and we multiplied with the energy unit (denominator on the l.h.s.). One readily confirms by performing the limit $\kappa \to 0$ that this result includes the Coulomb case. The model (\ref{eq:ns2}) differs from Refs~\cite{totsuji05c,henning06} by the additional correlation corrections which generalize the Coulomb expression (\ref{equation2}). This model was used in Ref.~\cite{baum07} and optimized to minimize the total energy. The detailed comparison to exact simulation data showed that, for Yukawa balls, the mean field model (neglecting the $\epsilon$ terms) performs rather poorly, but satisfactory agreement with an accuracy of the shell populations within $5\%$ can be achieved only by including correlation corrections and using different $\epsilon$-values for different shells. Yet a systematic derivation of the correlation corrections remains open.

\subsection{Continuum theory approach. Radial density profile of Yukawa balls}\label{ss:continuum}
An entirely different approach to the many-particle behavior of the Yukawa balls has been developed by {\sc Kraeft} and {\sc Bonitz} \cite{kraeft06b} and by 
{\sc Henning} et al. in Refs.~\cite{henning06,henning07}. It is based on a fluid-like statistical theory where the particles are treated as a continuum. This may seem far from reality but one can hope to describe several aspects of the Yukawa ball structure such as spatially averaged properties and achieve deeper physical insight. One such question is the mean radial density profile of the Yukawa balls. 
It is well known that, in a spherically symmetric parabolic potential, particles interacting via the Coulomb potential establish a radially constant-density profile (this follows from the fact that the potential profile inside a homgeneously charged sphere is a parabolic function of the distance from the center). However, in the case of Yukawa balls the pair interaction is screened, which leads to enhanced populations of inner shells with increaed $\kappa$ while the shell radii (in units of the mean interparticle distance within the shells) does not change with $\kappa$, cf. Sec.~\ref{ss:screening_struc}. From this we may expect that radial distribution of matter in the Yukawa balls will be inhomogeneous and the density has to decay towards the edge. This has indeed been observed experimentally \cite{kaeding08,block08} although the trend is weak, see Fig.~\ref{fig9}.

To analzye this question quantitatively we consider the model of a finite one-component plasma containing N identical particles described by the Hamiltonian (\ref{equation1}). The classical ground state energy follows from neglecting in Eq.~(\ref{equation1}) all particle momenta and can be written as a functional of the density \cite{henning06}
\begin{equation}\label{equation3}
	H_0[n] = E[n] = \int d^3r \,u(\textbf{r}),
\end{equation}
with the potential energy density 
\begin{equation}\label{equation4}
	u(\textbf{r}) = n(\textbf{r}) \left \{ \Phi(\textbf{r}) + 
\frac{N-1}{2N} \int d^3r_2 \,n( \textbf{r}_2 ) \frac{q^2}{4\pi\epsilon |{\bf r}-{\bf r}_2|}e^{-\kappa|{\bf r}-{\bf r}_2|}  \right \} + u_{corr},
\end{equation}
where the first term on the right is the mean-field contribution with the confinement potential $\Phi$, and $u_{corr}$ denotes the density of the correlation energy. The ground state density profile can be obtained from minimizing the total energy $E$, i.e. from the solution of the  variational problem $0 \stackrel{!}{=} \delta E [n] / \delta n(\textbf{r})$ under the constraint $\int d^3r \,n(\textbf{r})=N=$const.This problem can be solved for a general anisotropic confinement potential $\Phi({\bf r})$ in a mean-field approximation, i.e., $u_{corr}\equiv 0$, with the result \cite{henning06}: $4\pi q^2n({\bf r}) (N-1)/N=(\Delta -\kappa^2)\Phi({\bf r}) + \kappa^2\mu $, where $\mu$ is a Lagrange multiplier (chemical potential) assuring the normalization.
%
\begin{figure}[ht]
	\centering
	\includegraphics[width=7.5cm]{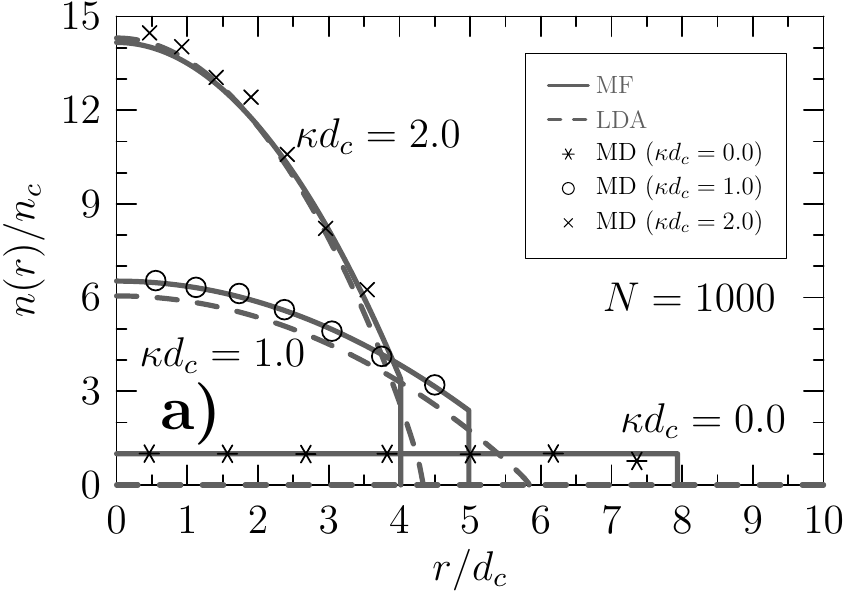}
	\includegraphics[width=7.5cm]{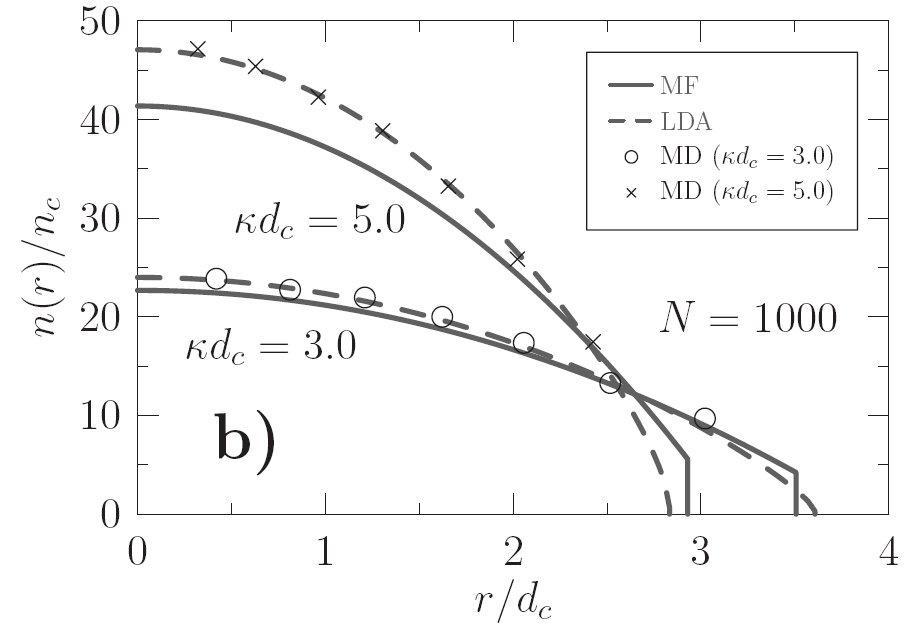}
	\caption{\label{fig5} Radial density profiles of a $3D$ plasma of $N=1000$ particles calculated with the  mean-field model (solid lines) and with the LDA including correlation contributions (dashed lines) for Coulomb and Yukawa interaction with four different screening parameters: $\kappa d_c = 1, 2, 3$, and $5$. Averaged shell densities of molecular dynamics simulations of a plasma crystal for the same parameters are shown by the symbols. $d_c$ is the two-particle equilibrium distance $r_0$, cf. Sec.~\ref{s:theory}}
\end{figure}

For a parabolic confinement, $\Phi(r) = \alpha / 2 r^2$, the ground state density profile is isotropic and parabolically decaying away from the trap center,
\begin{equation}\label{equation5}
	n(r) = \frac{\alpha N}{4 \pi (N-1) q^2} \left( c-\frac{\kappa^2 r^2}{2} \right) \Theta(R-r),
\qquad
c = 3+\frac{R^2 \kappa^2}{2} \frac{3+\kappa R}{1+\kappa R},
\end{equation}
and the density drops to zero at a finite radius $R(N,\kappa)$ which follows from the normalization
\begin{equation}\label{equation5c}
	-15 \frac{Q^2}{\alpha}(N-1)-15 \frac{Q^2}{\alpha} \kappa (N-1) R+15R^3+15\kappa R^4+6\kappa^2 R^5+\kappa^3 R^6 = 0.
\end{equation}
This equation has four complex and two real solutions, only one of which is non-negative, and thus constitutes the unique proper result entering equation (\ref{equation5}). This density is equivalent to a local force balance \cite{henning06,piel08_2}. 
Figure \ref{fig5} shows the density profiles derived from the mean field (MF) result (\ref{equation5}) (solid lines) and compares them to the averaged shell densities of MD simulations (symbols). In the case of weak screening (a) the MF result agrees perfectly with the exact MD results, while for increased screening discrepancies show up, especially in the trap center. 

The origin of these deviations is the neglect of correlations. An extension of the previous analysis to include correlation effects is difficult as the form of the pair correlation function is unknown. One way around this problem is to apply a local density approximation (LDA) \cite{totsuji01,henning07}. There one replaces the nonlocal terms within the energy density at point $r$ by local expressions using the known energy density of the homogeneous system \cite{totsuji05a}
\begin{equation}\label{equation5d}
	u^{corr}(n_0,\kappa) = -1.444 Q^2 n_0^{4/3} \exp \left[ -0.375 \kappa n_0^{-1/3} + 7.4 * 10^{-5} \left(\kappa n_0^{-1/3}\right)^4 \right],
\end{equation}
where $n_0$ is the corresponding density of the homogeneous system. Using this expression in equation (\ref{equation4}), leads to an improved equation for the density profile which has to be solved numerically. 
The resulting density profiles for a cluster with $1000$ particles are included in Fig.~\ref{fig5}. In the limit of strong screening, where the MF result failed to describe the simulation results, the LDA approximation with correlations is very accurate, see Fig.~\ref{fig5}b. In contrast, for weak and moderate screening, $\kappa r_0 \geq 2$, the LDA behaves very poorly. This is due to the long range of the interaction which cannot be accounted for within a local approximation. Thus, the mean-field model together with the presented local density approximation complement one another in the description of strongly correlated spatially confined one-component plasmas providing a complete theoretical description of the average ground state properties of Yukawa balls. An extension of these results to finite temperatures has been outlined in Ref.~\cite{wrighton09}

\section{Structural transitions}\label{s:transitions}
Up to this point we concentrated on the ground state properties of Yukawa balls, but as was shown, in 
Fig.~\ref{fig2} and table \ref{tab2} also other stationary states -- metastable states are frequently resolved in the simulations. 
Their energy is higher than that of the ground state but often only marginally. It may, therefore, be expected that these metastable states might be observable in the experiment and also transitions between stationary states might occur. We first analyze these questions theoretically and then compare with recent experimental results.

\begin{figure}[ht]
\centering
\includegraphics[width=7cm]{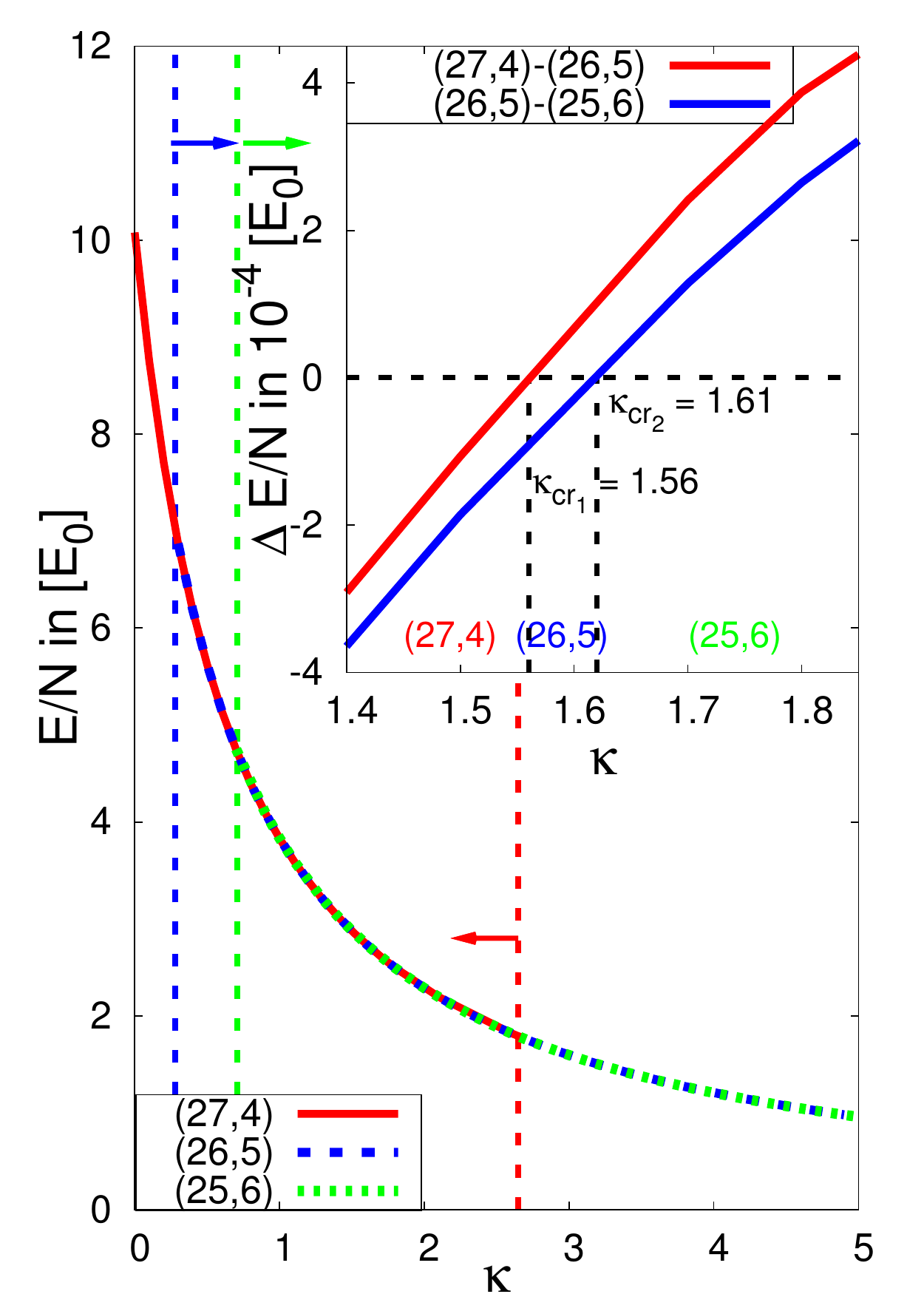}
\includegraphics[width=8cm]{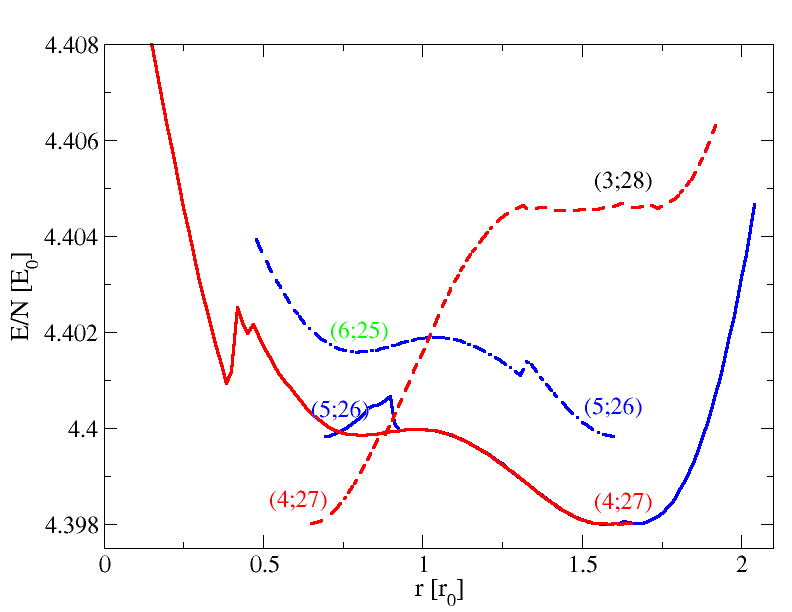}
\caption{\label{fig6} 
(colour online) {\bf Left}: Energy per particle of a Yukawa ball with $31$ particles vs. screenings strength. The ground state configuration is $(27,4)$ for $\kappa \le 1.56$, which is replaced by $(26,5)$ in the inverval $1.56 \le \kappa \le 1.61$, followed by $(25,6)$ for larger $\kappa$. Inset shows the energy differences of the configurations.
The vertical dashed lines with the arrows denote the beginning [blue for (26,5) and green for (25,6)] and the end [red for (27,4)] of occurance of these configurations in the simulations. 
{\bf Right}: Potential energy landscape measured by a MC simulation where one particle is moved, while all other particles are allowed to relaxate into a new minimum energy configuration ($N=31$, $\kappa=0.8$). Curves show the potential barriers for three transitions $(27,4) \rightarrow (26,5)$ (solid line), $(26,5) \rightarrow (25,6)$ (dashed) and $(27,4) \rightarrow (28,3)$. 
The peak in the solid curve around $0.75 < r_0 < 0.9$ occurs when the number of MC steps is reduced simulating incomplete relaxation. Interestingly this peak matches the return point in the experiment, cf. Fig.~\ref{fig9} (a).
}
\end{figure}
%

\subsection{Coexistence of stationary states. Potential barriers}\label{ss:screening}
As we have seen in Fig.~\ref{fig4}, upon increase of $\kappa$ the ground state configurations frequently change.
This requires that, around some critical value $\kappa_{cr}$, both, the old and new configuration are stationary states, corresponding to minima of the total potential energy $H_0$. When $\kappa$ departs from $\kappa_{cr}$ one 
of the configurations (the metastable state) will eventually disappear. This is illustrated for $N=31$ in the left part of Fig.~\ref{fig6}. The right (left) arrows indicate the range of $\kappa$ values where various metastable states appear (vanish).

If stationary states co-exist one could expect to observe transitions between them which may occur spontaneously, driven by thermal fluctuations. However, this does not only depend on the local minima but also on the potential landscape between them, in particular on the potential barrier.
The radial potential barriers can be ``measured'' in a computer simulation by moving, e.g. an outer particle to the inner shell, while allowing all other particles to relax. The results of such a MC simulation are shown in the right part of Fig.~\ref{fig6}. They indicate that, at the chosen screening, transitions $(6,25) \rightarrow (5,26)$ and $(5,26) \rightarrow (4,27)$ occur practically spontaneously whereas the opposite transitions are hampered by a significantly higher barrier. This asymmetry his even more pronounced for the transition between the configurations 
$(4,27)$ and $(3,28)$. This is nicely confirmed by the experiments \cite{block08} where a slow transition $(4,27) \rightarrow (5,26)$ could be observed followed by a rapid return to the ground state $(4,27)$, see Fig.~\ref{fig9}.




\subsection{Probability of occurance of metastable states}\label{ss:probability_ms}
\begin{figure}[ht]
\centering
\includegraphics[width=7.5cm]{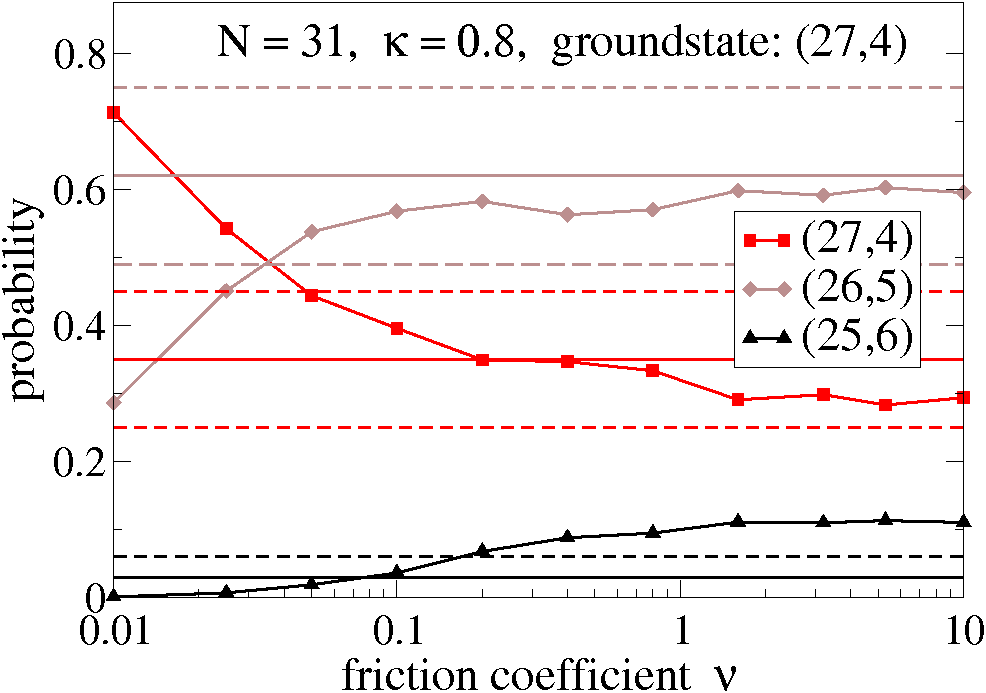}
\includegraphics[width=7.5cm]{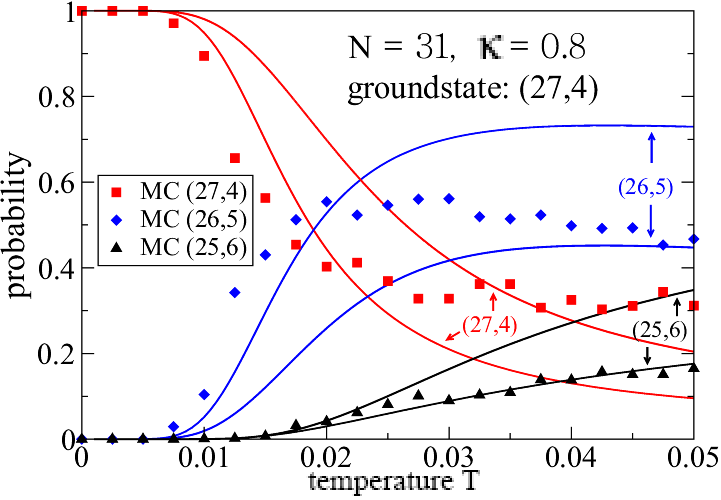}
\caption{\label{fig8} (color online) Experimental and theoretical probability of stationary states for a Yukawa ball with $N=31$.
{\bf Left:} Experimental results including mean probability (full lines) and standard deviations (dashes) corresponding to the states (from top to bottom) with $5, 4$ and $6$ particles on the inner shell compared to LMD results (lines with symbols) versus friction at $\kappa=0.8$.
{\bf Right:}
Analytical theory (lines) compared to MC results (symbols) as a function of temperature. The solid lines are obtained from Eq.~(\ref{equation14}) and the dashed lines neglect the statistical weight factor caused by the eigenfrequencies,  ($\Omega_s = 1$) for all states. Agreement with the experimental data is obtained for $T\sim 0.025$.}
\end{figure}

In the case of coexisting stationary states the question arises with what probability each of them will be observed when the Yukawa balls are formed. This subtle question could in fact be answered experimentally by performing a sequence of repeated measurements where the confinement was rapidly turned off and on again such that the plasma parameters were not altered \cite{kaeding08,block08}. The statistics was sufficient to extract reliable occurrence frequencies of different states which are shown in the left part of Fig.~\ref{fig8} by the horizontal solid straight lines (the dashed lines indicate the statistical error). The measurements proved the existence of three stationary states. Most suprisingly, the metastable configuration $(26,5)$ was found to occure with an almost two times higher probability than the ground state $(27,4)$ [$(62\pm 13)\%$ compared to $(35\pm 10)\%$]. We will now analyze these observations theoretically.

The occurence probability of stationary states can be described analytically by a harmonic approximation of the partition function \cite{kaehlert08}. There, the potential energy of a given state $s$ (ground or metastable state) is expanded around the local minimum $E^0_s$. With the assumption that the vibrational and the three rotational modes of the whole system are decoupled, the full energy of the state $s$ is, to second order in the displacements, 
$\xi_{s,i}=r_i-r_i^s$,
\begin{equation}\label{equation8}
E_s = E^0_s + \sum_{i=1}^f \left \{ \frac{p^2_{\xi_{s,i}}}{2m} + 
\frac{m}{2} \omega^2_{s,i} \xi^2_{s,i} \right \} + \sum_{i=3}^3 \frac{L^2_{s,i}}{2I_{s,i}}.
\end{equation}
The term in parentheses denotes the vibrational kinetic and potential energy of the $f=3N-3$ normal modes with the normal coordinates $\xi_i$. Even though the harmonic approximation is only applicable for low temperatures (or strong coupling) when the particles oscillate around the equilibrium positions with a small amplitude, it can be used to estimate phase transitions and compare the results with simulations and experimental data. With the factorized partition function
\begin{equation}
\label{equation9}
	Z_s = n_s Z_s^{int} Z_s^{vib} Z_s^{rot}, \quad 
\mbox{with the degeneracy factor} \quad	n_s = \frac{N!}{\prod_{i=1}^L N_i^s!},
\end{equation}
where $L$ is the number of shells and $N_i^s$ the occupation number of shell $i$ with $\sum_{i=1}^L N_i = N$. The degeneracy factor $n_s$ denotes the number of possibilities to form a configuration with shell occupation $(N_1,N_2, ..., N_L)$ from $N$ distinguishable particles. The internal partition function is given by
	$Z_s^{int} = e^{-\beta E_s^0}$,
and the vibrational and rotational partition functions are
\begin{equation}\label{equation12}
	Z_s^{vib}(T) = \left( \frac{k_BT}{\hbar \Omega_s} \right)^f, \qquad
	Z_s^{rot}(T) = \left( \frac{2 \pi k_B T \overline{I_s}}{\hbar^2} \right)^{3/2},
\end{equation}
where we introduced the geometric mean eigenfrequency $\Omega_s = (\prod_{i=1}^f \omega_{s,i})^{1/f}$ and the mean moment of inertia $\overline{I_s} = (I_{s,1} I_{s,2} I_{s,3})^{1/3}$. Then the occurence probability of the stationary state $s$ is obtained as
\begin{equation}\label{equation14}
	P_s = \frac{n_s Z_s}{Z} = \frac{n_s Z_s}{\sum_{s'=1}^M n_{s'} Z_{s'}}.
\end{equation}
The moments of inertia cancel to a good approximation for clusters with 27-40 particles, while for lower particle numbers they should be included. Since the total partition function $Z$ is not known, it is advantageous to compute instead the ratio of probabilites of two states $s$ and $s'$, using Eqs.~(\ref{equation12}) and~(\ref{equation14}) 
\begin{equation}\label{equation15}
\frac{P_s}{P_{s'}} = \frac{n_s}{n_{s'}} \left( \frac{\Omega_{s'}}{\Omega_s} \right)^f \left( \frac{\overline{I_s}}{\overline{I_{s'}}} \right)^{3/2}  e^{-\beta (E_s^0-E_{s'}^0)} \approx \frac{n_s}{n_{s'}} \left( \frac{\Omega_{s'}}{\Omega_s} \right)^f e^{-\beta (E_s^0 - E_{s'}^0)}.
\end{equation}

Thus the probability ratio of two states depends on three factors: their energy difference $E_s^0 - E_{s'}^0$, the ratio of degeneracy factors $n_s/n_{s'}$ and the ratio of mean eigenfrequencies $\Omega_{s'}/\Omega_s$.
The Boltzmann factor $e^{-\beta (E_s^0 - E_{s'}^0)}$ gives preference to states with a low energy. For low temperatures it will be the dominant factor but it becomes less important for higher temperatures when $k_B T >> E_{s'}^0 - E_s^0$ and $e^{-\beta (E_s^0 - E_{s'}^0)} \approx 1$.
The degeneracy factor assigns large statistical weight to states with more particles on inner shells. For example, for $N=31$ the ratio of the factors for two configurations with, respectively, $4$ and $5$ particles on the inner shell is $n_{(27,4)}/n_{(26,5)} = (26!5!)/(27!4!) = 5/27$. 
The energy differences are computed by MD simulations and, as an exmmple for $N=31$ and $\kappa = 1.4$, are given in table \ref{tab2}. The effect of the mean eigenfrequency, i.e. the effect of the local curvature of the potential surface cannot be that easily predicted but the ratio of the eigenfrequencies are in the range of $0.8 - 30$ and can therefore strongly influence the ratio of the probabilty of states. 
The analytical results can be compared with MC simulation results \cite{kaehlert08} which are practically exact. In particular, they include all anharmonic corrections. For a fixed temperature the occurrence probability of each stationary state is recorded. Both results are shown in Fig.~\ref{fig8} as a function of temperature, for the example $N=31$. Evidently, the analytical results are accurate at temperatures below $k_B T < 0.02$. Above this value anharmonic corrections become important. For other particle numbers and screening parameters simliar results are obtained. 

Let us now compare to the experimental results, cf. left part of Fig.~\ref{fig8}. These data were obtained at room temperature, corresponding to approximately $T=0.0015$. At this temperature the MC and analytical results both are in agreement with each other (right figure) predict practically $100$ percent occurence of the ground state $(27,4)$ which is in striking contrast to the experiment. To resolve this contradiction, independent Langevin dynamics simulations have been performed \cite{block08,kaehlert08} which model the experimental situation of turning on and off the confinement by including a time-dependent potential $U^{\rm ext}$ in Eq.~(\ref{equation1}) and which are included in the left figure part. Using a  screening parameter which corresponds to the experimental parameters (the same as in the MC simulations) 
one finds a strong dependence on the chosen values of the friction coefficient $\nu$. For small $\nu$ the simulation always favor the ground state which is easily understood: with small friction the system is cooled very slowly and has enough time to find the lowest potential minimum (this is nothing but simulated annealing which was applied in Sec.~\ref{ss:groundstate} to obtain the ground states). In contrast, with large damping the particles rapidly loose their kinetic energy even before settling in a potential minimum. The experimental damping parameters are well above $\nu=2$ where Fig.~\ref{fig8} indicates very good agreement between LMD and experiment. On the other hand, Monte Carlo and analytical theory are based on a hamiltonian model, cf. Sec.~\ref{ss:simulations}, and do not include damping effects. Thus, a direct comparison with the experimental probabilities is not possible. However, as was demonstrated in Ref.~\cite{kaehlert08} due to the large friction in the experiments the plasma particles rapidly relax towards a Maxwellian distribution already at temperatures of about $T=0.05$. During further cooling the system remains in equilibrium, only its temperature is being gradually reduced until it reaches $T\sim 0.002$. The ``decision'' about what stationary state will be reached is ``made'' at an early stage, before the Yukawa ball freezes which occurs around $T\sim 0.02$. In fact, Fig.~\ref{fig8} shows that the MC simulations at $T\approx 0.025$ agree very well with the experiment and the LMD \cite{kaehlert08}.

\subsection{Structural transitions in the experiment}\label{ss:phase_trans_exp}
\begin{figure}[ht]
\centering
\includegraphics[width=9cm]{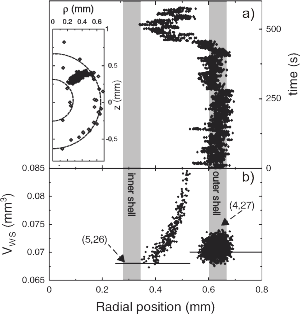}
\caption{\label{fig9} 
(a) Radial component of the trajectory of one particle leaving the outer shell of a Yukawa ball with $N=31$, the inset shows the trajectory in the $\rho$-$z$ plane (black symbols). (b) Average Wigner-Seitz cell volume per particle, $V_{WS}$, for all particles belonging to the inner shell as a function of the radial position of the traveling particle. For $R=0.55 mm$, the Wigner-Seitz cell of the traveling particle is included in the average. The solid lines are the average $V_{WS}$ of the configurations $(4,27)$ and $(5,26)$ obtained by repeating the experiment and averaging.}
\end{figure}

Transitions between two stationary configurations which were discussed above could, in fact, be observed 
 experimentally \cite{block08}. Figure~\ref{fig9}(a) shows the motion of a single particle of a Yukawa ball with $31$ particles over a period of ten minutes. It can be seen that the particle moves from the outer shell inwards and returns to the outer shell within about $200$ seconds. This transition is temperature induced (it occurs spontaneously) and is observed in other Yukawa balls as well \cite{block08}.
This behavior is well explained by the MC simulation of the total potential profile [see the discussion in Sec.~\ref{ss:screening}] which are shown in the right part of Fig.~\ref{fig6}. The potential barrier for moving a particle inwards from $(27,4)$ to $(26,5)$ [lowest curve] shows where the particle returns; interestingly the sharp increase at $r_0 \approx 0.85$ is matching the return point of $r=0.39mm$ in the experiment, cf. Fig.~\ref{fig9}. 
This peak corresponds to a MC run with fewer steps where the other particles have not enough time to fully relax in response to the moving paricle -- this nonequilibrium situation is apparantly observed in the experiment.

\subsection{Finite temperature effects. Solid-liquid phase transition}\label{ss:finite_t}
So far we have considered only the low-temperature behavior of the Yukawa balls where the system was in the ground state or low-lying metastable state. Finite thermal energy caused only occasional transitions between stationary states. This behavior is typical for matter at very strong coupling characterized by very low kinetic energy, typical for a Wigner crystal. This physical situation can be quantified by introducing
a {\em coupling parameter}, which is defined as the ratio of the mean interaction energy to kinetic energy. For a classical macroscopic Coulomb system in thermal equilibrium the definition is
\begin{equation}\label{equationtemp1}
	\Gamma_C = \frac{q^2}{4\pi \epsilon a k_B T},
\end{equation}
where $a$ is the mean interparticle distance. The transition from strong coupling where the system is in a crystalline state to a liquid occurs at critical value $\Gamma_{cr}$ which is of the order $175$ ($137$) in $3D$ ($2D$). The transition to a gas-like state occurs around $\Gamma_C=1$,  (for generalizations to quantum systems, see Ref.~\cite{bonitz08}).

Here we are concerned with systems with Yukawa interaction and have to generalize the coupling parameter. The straighforwad way to do this would be to replace the Coulomb energy of a pair of particles by the mean Yukawa energy and $\Gamma_C \rightarrow  e^{-\kappa a}\cdot \Gamma_C$, but this makes it impossible to systematically define a value for the melting transition in a Yukawa system, so for the phase diagram \cite{hamaguchi97} one has to use other quantities. On the other hand, it is not the total potential energy which governs the melting transition, but the process is driven by the excitation energy of particles above the potential minimum which leads to increasing fluctuations of the particles around their stationary positions. These fluctuations are defined by the local curvature of the potential \cite{bonitz06,bonitz08}, i.e. its second derivative in the minimum. This allows to define a proper effective Yukawa coupling parameter \cite{vaulina00},
\begin{equation}
\label{equationtemp3}
\Gamma (\kappa) = \Gamma_C \frac{e^{\kappa a}}{1+\kappa a + (\kappa a)^2/2},
\end{equation}
which allows to qualitatively correctly predict the melting point $\Gamma_{cr}(\kappa)$ for any $\kappa$, based on the known critical value for a Coulomb system, i.e. by using, in Eq.~(\ref{equationtemp3}), $\Gamma_C=\Gamma_{cr}$.

However, for the analysis of melting in Yukawa balls one has to take into account the peculiarity of small systems.
{\em Finite systems} behave fundamentally different to macrososcopic systems: there exist no strict phase transitions. Nevertheless, a similar transition or {\em crossover} from solid-like to liquid-like behavior is well known and has been studied as a function of particle number e.g. by {\sc Schiffer} \cite{schiffer02}. For small $N$ the melting temperatures are typically higher than in the macroscopic limit and the crossover proceeds over a finite range of coupling parameters (temperatures) making it often difficult to identify a melting point, see below. 
Furthermore, finite systems are strongly influenced by the symmetry of the confinement potential, and the shell structure of the Coulomb balls affects the melting behavior.

The effect of screening on the stability and melting of Yukawa balls was analyzed in Ref.~\cite{bonitz06}, 
where a simple analytical estimate can be observed for the example $N=2$. Expanding the total potential energy to second order around the ground state, as in the macroscopic system, a formula analogous to (\ref{equationtemp3}) is obtained
\begin{equation}\label{equationtemp2}
	\Gamma (\kappa,N=2) = \Gamma_C(N=2) \frac{e^{\kappa r_{0Y}}}{1+\kappa r_{0Y}+\kappa^2 r_{0Y}^2/3}.
\end{equation}
Here $r_{0Y}(\kappa)$ is the two-particle distance for Yukawa interaction which is obtained by inverting 
the equation
\begin{equation}\label{equationtemp2b}
	\frac{e^{\kappa r_{0Y}} r_{0Y}^3}{1+\kappa r_{0Y}} = \frac{2\omega^2q^2}{4\pi \epsilon m} = r_0^3(r_{0Y}),
\end{equation}
where $r_0$ is the equilibrium distance of two particles with Coulomb interaction, see Sec.~\ref{s:theory}.
As in the macroscopic case, screening always destabilizes the crystal.

\begin{figure}[ht]
\begin{minipage}[c]{6.3cm}
\includegraphics[width=3.7cm,angle=-90]{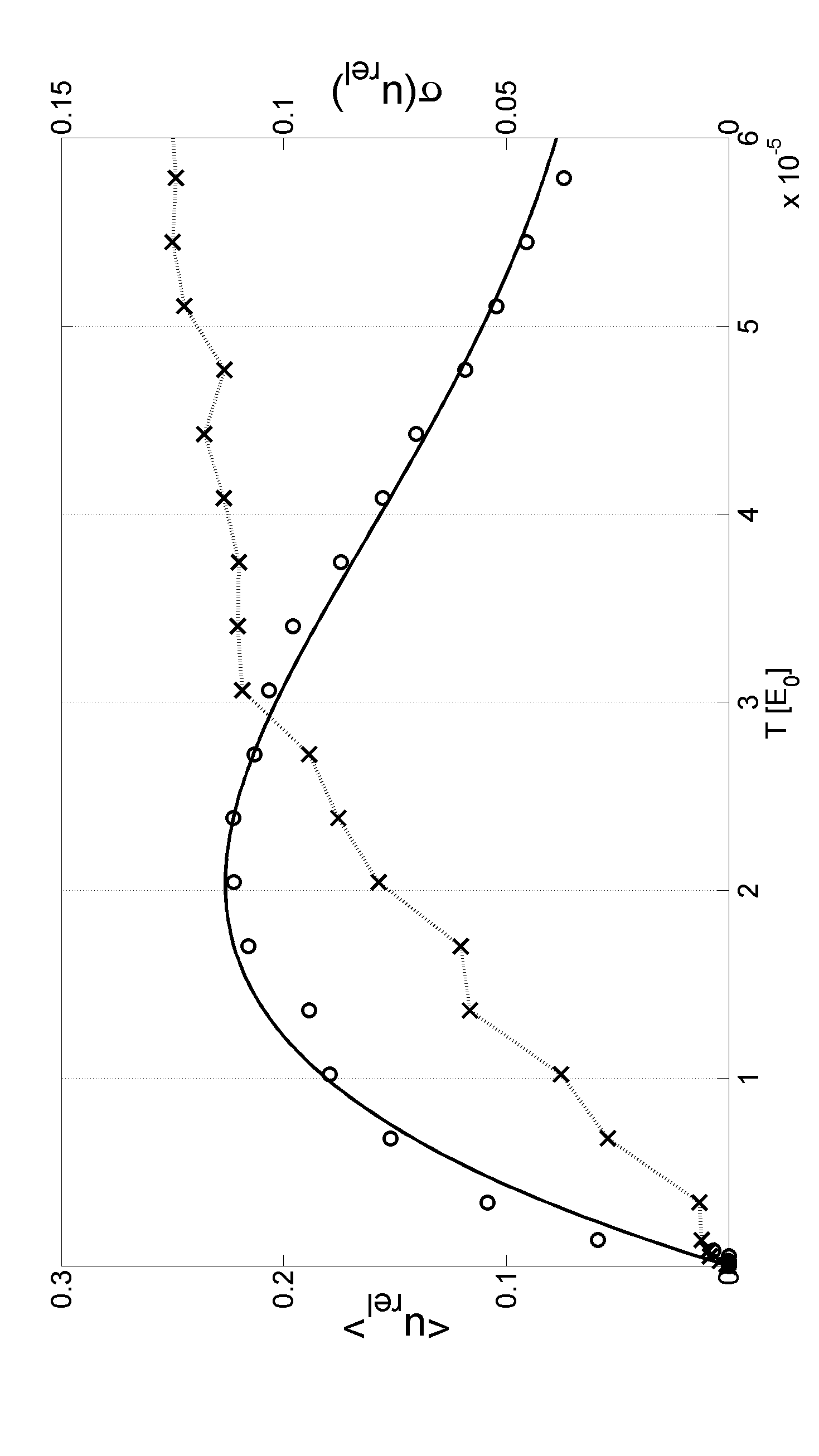}
\end{minipage}
\vspace{0.1cm}
\begin{minipage}[c]{9cm}
\includegraphics[width=6cm,angle=-90]{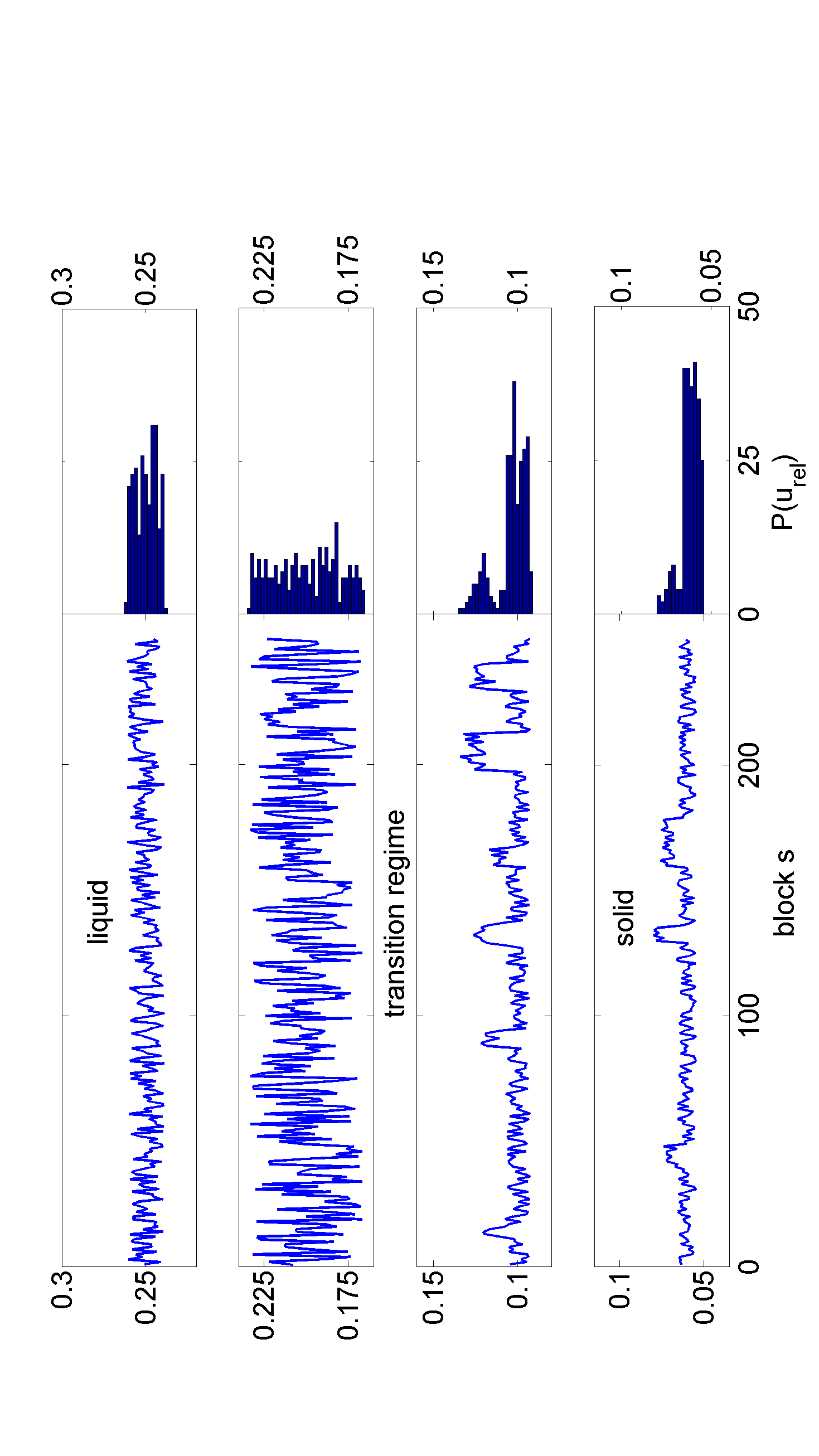}
\end{minipage}
\caption{\label{fig7} Mean distance fluctuations (\ref{equation7}), crosses and their variance (\ref{equation6}), circles, for a Yukawa ball with $\kappa=0.4$ in the vicinity of the melting point $T_{cr}$. {\bf Left}: Temperature dependence of $u_{rel}$ and $\sigma_{u_{rel}}$ for a Yukawa ball with $N=40$. {\bf Right:} Block averaged interparticle distance fluctuations $u_{rel}(s)$ vs block number s during an MC simulation and their accumulated probability $P$ (rightmost figure) for $N = 40$ and four temperatures (from bottom to top): below, close to, at and above $T_{cr}$. Block length $M = 1000$.}
\end{figure}

Another quantity to locate the melting point are the {\em mean position fluctuations} of particles which are known to be small (large) in the solid (liquid) phase, independently of the interaction and system size \cite{bonitz08} which is commonly known as Lindemann criterion. However, this parameter diverges in $2D$ systems, and an improved quantity are the mean {\em relative interparticle distance fluctuations}, i.e. the fluctuations normalized to the mean distance
\begin{equation}\label{equation7}
u_{rel} = \frac{2}{N(N-1)} \sum_{1\leq i \leq j}^N \sqrt{\frac{\langle r_{ij}^2 \rangle}{\langle r_{ij} \rangle^2}-1}.
\end{equation}
This quantity has been successully applied to many finite systems and also to Yukawa balls 
\cite{golubnychiy06,apolinario07,baum07,apolinario_phd}, an example is shown in the left part of Fig.~\ref{fig7}. 
One clearly sees the increase of the fluctuations typical for a melting process. It is also evident that this increase extends over more than one order of magnitude in the temperature. 
A rather detailed analysis of melting in Coulomb balls was performed by {\sc Apolinario} \cite{apolinario_phd} who observed that the melting often proceeds in several steps. 

Finally, in a recent study \cite{boening08} it was observed that the quantity $u_{rel}$ has a number of difficulties in small systems -- the result strongly depends on the way of calculation. Instead it was suggested to use another quantity -- the variance of the block averaged interparticle distance fluctuations (VIDF)
\begin{equation}\label{equation6}
\sigma_{u_{rel}} = \frac{1}{K} \sum_{s=1}^K \sqrt{\langle u_{rel}^2(s) \rangle - \langle u_{rel}(s)\rangle^2}.
\end{equation}
This parameter measures the width of the probability distribution $P(u_{rel})$ which has a maximum in the vicinity of the phase transition. In the solid phase the particles show only small distance fluctuations, and in the liquid phase all particles strongly fluctuate and $P(u_{rel})$ has a peak centered around a large value, yet the width of 
the peak is again small. However, in the vicinity of the phase transition, the particles may be in one state for some time and frequently make transition over the potential barrier  to another stable state  and so on. 
This causes a broad peak of $P$, as can be seen in the rightmost part of Fig.~\ref{fig7}.
%
%

An experimental analysis of the melting behavior is difficult because it require a systematic scan of the plasma parameters in a sufficiently broad range around the expected melting point. To do this without changing the screening is difficult and requires future advances in the handling of the Yukawa balls.

\subsection{Further results on Yukawa balls}\label{ss:further}
In this review we had to concentrate on a few properties of Yukawa balls leaving out many other interesting results some of which are mentioned in the following.

The normal mode spectrum of Yukawa balls which was mentioned in Sec.~\ref{ss:probability_ms} has been analyzed in great detail. Among the most interesting collective excitations of these finite systems is the breathing mode (uniform radial expansion and contraction). Recently it could be shown \cite{henning08} that Yukawa balls do not possess such a mode. Instead they can exhibit simultaneously several monopole oscillations (periodic increase and reduction of the mean radius). Also, first experimental observations of the normal modes of small Yukawa balls have been performed \cite{henning09}.

The confinement potential in the experiments is certainly never perfectly isotropic, so the analysis of anisotropy effects is of interest. Ground and metastable states in anisotropic parabolic confinements have been analyzed as a function of the anisotropy by {\sc Apolinario} et al. \cite{apolinario07,apolinario_phd}. They also analyzed the normal modes of these systems. The results are of interest also for experiments where the confinement is made anisotropic on purpose. 

Another interesting theoretical investigation has been performed by {\sc Psakhie} et al. They  performed MD simulations for Yukawa balls containing two sorts of particles having different masses \cite{abdrashitov08}. In their model the confinement forces, the termophoretic force, gravity, electric field force and ion drag have been explicitely included. Increasing the mass difference they found that beyond a critical mass asymmetry the Yukawa ball may split in two balls containing different particles which are vertically separated.

Recently, first simulations of the short-time dynamics of Yukawa balls have been performed by {\sc K\"ahlert} et al. \cite{kaehlert08}. Starting from a random initial state and rapidly turning on the confinement showed a very interesting nonequilibrium behavior with a non-monotonic kinetic energy increase. This behavior is very different from the ultrafast dynamics of weakly coupled plasmas and is a very interesting subject of further research.
%

\section{Discussion and Outlook}\label{s:dis}
In this review, and overview on the recently discovered spherical crystals in dusty plasmas -- Yukawa balls -- has been given. The experimental setup and the diagnostic methods have been briefly discussed. The main attention was then devoted to the theoretical modeling and to computer simulatinos of Yukawa balls, when possible, aiming at a detailed comparison to the experimental data. 
This close connection between experiment and theory has lead to substantial advances in our understanding of finite Yukawa crystals. It was shown that the Yukawa balls produced in the Kiel and Greifswald labs are well described
by an isotropic statically screened pair potential and that wake effects are apparantly not of importance. This situation may change when larger balls are studied which is an interesting question for further research.

It was shown that screening has a crucial effect on the internal structure of Yukawa balls, including the 
shell populations, the ground and metastable states. Comparisons between measurement and simulations allowed to accurately determine the value of the screening parameter. Besides the shell structure, also the mean distribution of charge throughout the Yukawa balls could be explained. This was achieved by an analytical theory based on a fluid-like description. It could be shown that, with increased screening, the average density is more and more concentrated in the central region of the cluster and rapidly decays towards the edge. This is an important difference to Coulomb systems, including spherical ion crystals, which exhibit a constant radial density profile. 

Not only the ground state of the Yukawa balls could be observed. The experiments were able to reliably detect metastable states and to reveal the probability of their occurence. These measurements could be well explained by theory and simulations where it became clear that the (in some cases) dominant role of metastable states -- despite relatively low temperature -- is due to the large friction experienced by the dust particles in the plasma. 
Besides the static properties also several dynamic aspects are now well understood. The experiments could demonstrate that spontaneous transitions between various stationary states occur which is well explained by a theoretical analysis of the potential energy landscape, in particular from the energy barriers between local potential minima. The theory further predicted a rather interesting melting behavior of the Yukawa balls upon further temperature increase. It will be of high interest to verify these predictions in future experiments. 

At the same time many fundamental theoretical questions remain to be answered. In the shell strucutre of the Yukawa balls correlation effects play a crucial role. This became clear by studying greatly simplified shell models. These effect were also important to explain the density profile of Yukawa balls at large screening using a local density approximation. Yet no analytical results for the pair correlations are known and no derivation of these effects from statistical mechanics has been given. The excellent experimental accuracy gives a unique opportunity to advance and verify modern theoretical approaches in strongly coupled plasmas.

Finally, questions of further interest in the field of Yukawa balls include their normal mode spectrum, 
and their interaction with external electric and magnetic fields. This will require to take into account the modification of the pair interaction. Here also wake effects could play a more important role.
Last but not least it is becoming possible to advance to an analysis of the short-time behavior Yukawa balls. Excerting a fast excitation, e.g. by laser manipulation, the time-dependent response of these strongly correlated systems can be systematically studied.

\section*{Acknowledgements}
We acknowledge stimulating discussions with J.W. Dufty, Ch. Henning, H. K\"ahlert, P. Ludwig, A. Melzer and A. Piel.
This work is supported by the Deutsche Forschungsgemeinschaft via SFB-TR 24 Projects A3, A5, and A7. M.B. acknowledges hospitality of the Physics Department of the University of Florida.

\end{document}